\documentclass[aps,prl,preprint,superscriptaddress]{revtex4-2}
\usepackage{CJK}
\usepackage{natbib}
\usepackage{graphicx}
\usepackage{amsmath}
\usepackage{amssymb}
\UseRawInputEncoding
\begin{document}
\begin{CJK*}{GB}{} % Use default fonts from CJK (see below)

%\preprint{}

\title{Boosting thermal conductivity by surface plasmon polaritons propagating along a thin Ti film}


\author{Dong-min Kim}
\affiliation{Department of Mechanical Engineering, Korea Advanced Institute of Science and Technology, Daejeon 34141, South Korea}
\affiliation{Center for Extreme Thermal Physics and Manufacturing, Korea Advanced Institute of Science and Technology, Daejeon 34141, South Korea}

\author{Sinwoo Choi}
\affiliation{Department of Mechanical Engineering, Korea Advanced Institute of Science and Technology, Daejeon 34141, South Korea}
\affiliation{Center for Extreme Thermal Physics and Manufacturing, Korea Advanced Institute of Science and Technology, Daejeon 34141, South Korea}

\author{Jungwan Cho}
\affiliation{School of Mechanical Engineering, Sungkyunkwan University, Suwon 16419, South Korea}

\author{Mikyung Lim}
\email{mklim@kimm.re.kr}
\affiliation{Nano-Convergence Manufacturing Systems Research Division, Korea Institute of Machinery and Materials, Daejeon 34103, South Korea}

\author{Bong Jae Lee}
\email{bongjae.lee@kaist.ac.kr}
\affiliation{Department of Mechanical Engineering, Korea Advanced Institute of Science and Technology, Daejeon 34141, South Korea}
\affiliation{Center for Extreme Thermal Physics and Manufacturing, Korea Advanced Institute of Science and Technology, Daejeon 34141, South Korea}

%\author{}
%\thanks{}
%\affiliation{}

\maketitle
\end{CJK*}

\date{\today}

\begin{abstract}
%600 characters
We experimentally demonstrate a boosted in-plane thermal conduction by surface plasmon polaritons (SPPs) propagating along a thin Ti film on a glass substrate. Owing to a lossy nature of metal, SPPs can propagate over centimeter-scale distance even with a supported metal film, and resulting ballistic heat conduction can be quantitatively validated. Further, for a 100-nm-thick Ti film on glass substrate, a significant enhancement of in-plane thermal conductivity compared to bulk value ($\sim 35\%$) is experimentally shown. This study will provide a new avenue to employ SPPs for heat dissipation along a supported thin film, which can be readily applied to mitigate hot-spot issues in microelectronics.

%original
% Recent experiments have revealed an enhanced in-plane thermal conduction along a polar dielectric film by a novel heat transfer channel exploiting surface phonon polaritons. However, the increase of thermal conductivity is substantial only for a suspended membrane, making quantitative analysis as well as practical application difficult. Here, we experimentally demonstrate a boosted in-plane thermal conduction by surface plasmon polaritons (SPPs) propagating along a thin Ti film on a glass substrate. Owing to a lossy nature of metal, SPPs can propagate over centimeter-scale distance even with a substrate, and resulting ballistic heat conduction can be quantitatively validated. Further, for a 100-nm-thick Ti film on glass substrate, a significant enhancement of in-plane thermal conductivity compared to bulk value ($\sim 35\%$) is experimentally shown. This study will provide a new avenue to employ SPPs for heat dissipation along a supported thin film, which can be readily applied to mitigate hot-spot issues in microelectronics.
\end{abstract}

%\keywords{}

\maketitle

\section{}

Nanoscale thin films cause a classical size effect on thermal conductivity as their thicknesses become smaller than the mean free paths of phonons or electrons \cite{qiu1993size, zhang2007nano}.Decreased thermal conductivity impedes heat spreading from hot spots within the devices, deteriorating their performance. As a solution, surface electromagnetic waves have been spotlighted as auxiliary heat carriers due to their orders longer propagation length than primary heat carriers (i.e., phonons and electrons) \cite{raether1988surface, mulet2001nanoscale, chen2005surface, ordonez2013anomalous, gluchko2017thermal, burke1986surface, tranchant2019two}, which can compensate for the size effect of film thermal conductivity.

Surface phonon polaritons (SPhPs), surface electromagnetic waves coupled with optical phonons, can carry heat and thus enhance the dielectric film thermal conductivity \cite{joulain2005surface, chen2010heat, baffou2010mapping}. Over the past few decades, thermal transport via SPhPs has been thoroughly studied for suspended dielectric membranes \cite{chen2005surface, ordonez2013anomalous, tranchant2019two, wu2020enhanced}. Chen \textit{et al.} proposed first that SPhPs can enhance the in-plane thermal conductivity of a 40-nm-thick-SiO${_2}$-suspended-membrane by more than 100\% of its bulk value \cite{chen2005surface}. SPhPs have an orders of magnitude longer propagation length ($\lambda$) than acoustic phonons in a SiO${_2}$ membrane \cite{chen2007measurement, gluchko2017thermal}. Subsequently, the SPhP thermal conductivity of a suspended dielectric membrane was experimentally demonstrated \cite{tranchant2019two, wu2020enhanced}. Tranchant \textit{et al.} measured the in-plane thermal conductivity of the suspended SiO${_2}$ membrane with respect to its thickness and width by using the 3$\omega$ method \cite{tranchant2019two}. Due to relatively large measurement errors, their results could not conclusively reveal the significant enhancement of SiO${_2}$ membrane thermal conductivity. Later, Wu \textit{et al.} used the micro time-domain thermoreflectance method to observe the SPhP thermal conductivity of suspended SiN membranes depending on their temperature \cite{wu2020enhanced}. However, their results could not reveal a strong link between the propagation lengths of the SPhPs and the resulting thermal conductivity. Such limitations in early measurements are mainly due to the limitations of the fabrication of nanoscale suspended membranes with large surface area. Although a supported thin film structure is preferred in terms of both experimental validation and real-world application, for a polar dielectric thin film on a substrate, a surface electromagnetic wave can only be supported when a film thickness is greater than the cutoff thickness ${d_c}$\cite{ordonez2013anomalous, yang1991long}, which causes significantly reduced thermal conduction by surface waves \cite{ordonez2013anomalous,lim2019thermal}. Consequently, there has been no experimental demonstration of surface-wave-enhanced thermal conductivity on a supported structure so far.

Surface plasmon polartions (SPPs) are other forms of surface electromagnetic waves coupled to free electrons in metals, which can be thermally excited and function as heat carriers.  Thermally excited SPPs in lossy metals or heavily-doped semiconductors exist in broad spectral regions and have been successfully employed to tune the near-field thermal radiation \cite{basu2011near, biehs2012hyperbolic, messina2013graphene,liu2014near, jin2016hyperbolic, lim2018optimization, lim2018tailoring}. In fact, SPPs supported at a lossy metal/dielectric interface can have propagation lengths of several millimeters or more in the mid-infrared regime \cite{burke1986surface}. Furthermore, long-range SPPs can propagate over centimeter-scale distances in a \textquoteleft thin\textquoteright \text{ }metal film \cite{berini2009long}. Therefore, SPPs supported in the thin lossy metal film can lead to a significant enhancement of its thermal conductivity. In this work, we experimentally demonstrate for the first time that SPPs on a thin lossy metallic film can also be exploited to enhance its in-plane thermal conductivity even in a supported configuration. We deposit a thin Ti film on a glass substrate with circular patterns, varying its radius. The size effect of SPP-enhanced thermal conductivity is clearly demonstrated by limiting the long propagation of SPPs (a few centimeters) to a smaller radius. Also, the effect of Ti-film thickness is quantitatively analyzed by varying its thickness (e.g., 100 nm, 300 nm, or 1000 nm). The measured in-plane thermal conductivity of Ti films is compared with a theoretical prediction based on the Boltzmann transport equation to reveal the relationship between the SPP propagation length and thermal conductivity.

%%%%%%%%%%
\begin{figure}[t!]%
\centering
\includegraphics[width=1\textwidth]{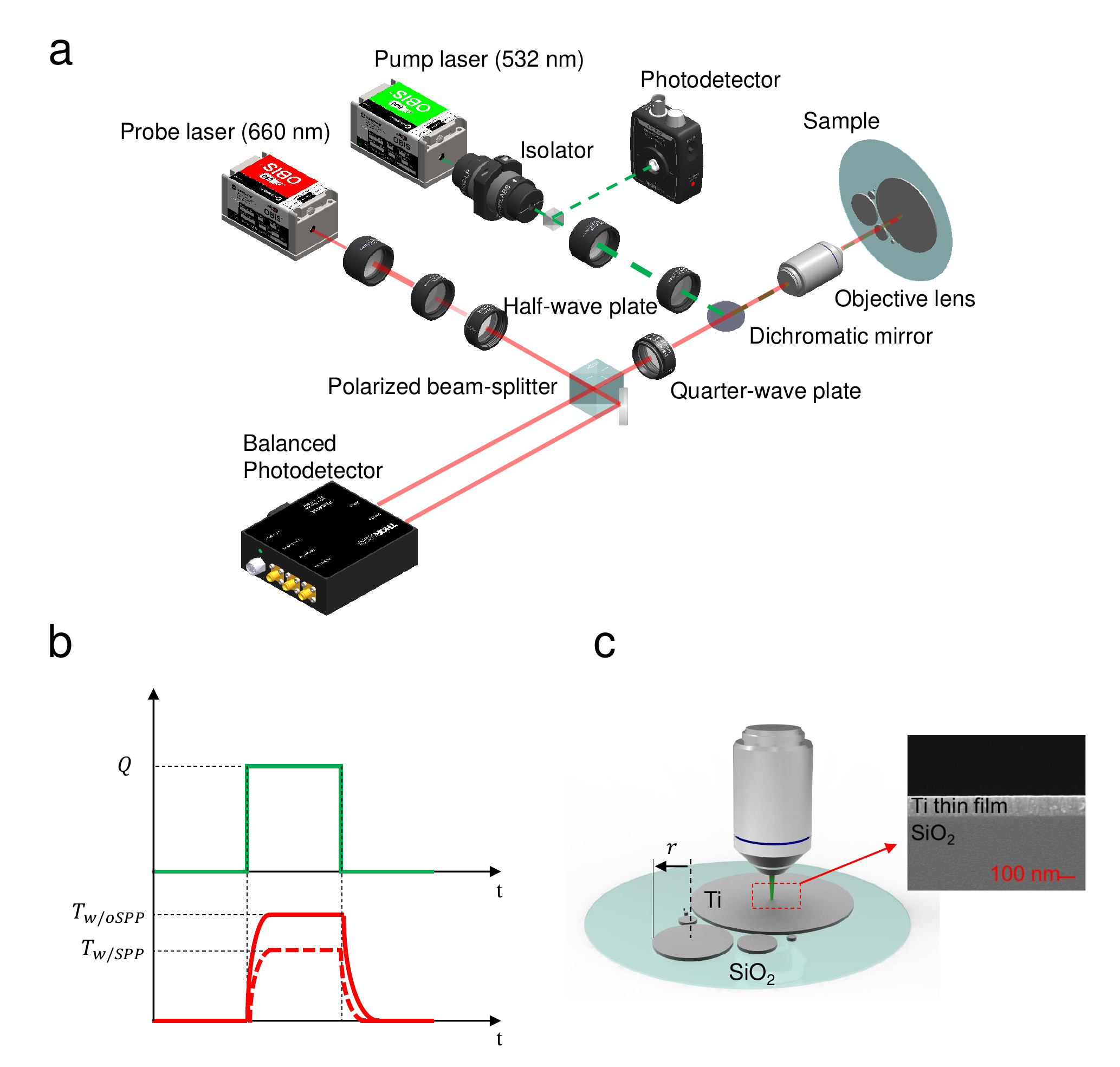}
\caption{Experimental setup for measuring in-plane thermal conductivity of Ti films. (a) Schematic of a custom-built steady-state thermoreflectance setup. (b) Temperature rise at the sample surface corresponding to the applied laser irradiation with power $Q$. The rise of the temperature of the heated region drops if there exists a ballistic thermal transport via SPPs supported along a thin Ti film. (c) Schematic of a set of measurement samples. Ti films were patterned in circular shapes with different radius values $r$ on a single glass (i.e., amorphous SiO$_2$) substrate. Inset shows the cross-section of the deposited Ti film with thickness of 100 nm. }\label{fig1}
\end{figure}
%%%%%%%

We employed a steady-state thermoreflectance method (SSTR) \cite{braun2019steady} to measure the in-plane thermal conductivity of Ti films, as shown in Fig. \ref{fig1}a. Section 1 of Supplemental Material describes our SSTR setup and validation. The SSTR is a variation of the frequency-domain thermoreflectance method (FDTR), which heats the sample near steady-state by reducing the modulation frequency of the pump laser. Due to its long thermal penetration depth, the SSTR has good measurement sensitivity on the in-plane thermal conductivity of thin films (Fig.\ S2). The in-plane thermal conductivity of thin films can then be obtained by fitting the measured surface temperature of the heated spot with the theoretical prediction derived from the two-dimensional (2-D) heat diffusion model \cite{braun2017upper}(Section 3 of Supplemental Material). We used the Ti film as a transducer considering its optical property \cite{abad2017non}; that is, there is no additional metal layer deposited for the measurement. Because the thermal conductivity of a glass substrate (i.e., $k_s=1.35$ W/m$\cdot$K; Table\ S2) is much lower than that of Ti ($\sim$10 W/m$\cdot$K), heat is mainly transferred through the Ti film like a suspended structure, increasing the measurement sensitivity of the in-plane thermal conductivity of the Ti film. Since the SSTR cannot distinguish the SPP-enhanced thermal conductivity from the diffusive electronic counterpart of Ti film, we experimentally investigated the thermal conductivity via SPPs with the SPP-induced temperature drop of the Ti film by comparing the temperature rise of the Ti film surface with and without SPP excitations under the same amount of laser irradiation, as illustrated in Fig.\ \ref{fig1}b. 

For measurements, a thin Ti film was first e-beam deposited on a glass substrate with various thicknesses and then circularly patterned by lift-off process. The Ti films were patterned with radius ranging from 200 $\mu$m to 28 mm. As shown in the inset of Fig.\ \ref{fig1}c, the Ti films were uniformly deposited. Due to boundary scattering effect resulting from the radius of Ti-film patterns, the effective propagation length ($\Lambda_{eff}$) of SPPs was determined by Matthiessen\textquoteright s rule (i.e., $\Lambda_{eff}^{-1}=\Lambda^{-1}+{r}^{-1}$ , where $r$ is the radius of the Ti-film pattern). Also, the spectral reflectance of the Ti film at wavelengths from 500 nm to 700 nm, which includes the pumping wavelength (i.e., $\lambda=532$ nm) and the probing wavelength (i.e., $\lambda=660$ nm), was measured with a UV-VIS spectrometer (UV-3600 Plus, Shimadzu) and agreed excellently with the theoretical prediction derived from the optical property of Ti using the Drude model \cite{ordal1988optical}(see Fig.\ S3). Thus, the Drude model is used when calculating the dispersion of SPPs for a thin Ti film deposited on a glass substrate.

%%%%%%%%%%
\begin{figure}[!b]%
\centering
\includegraphics[width=1\textwidth]{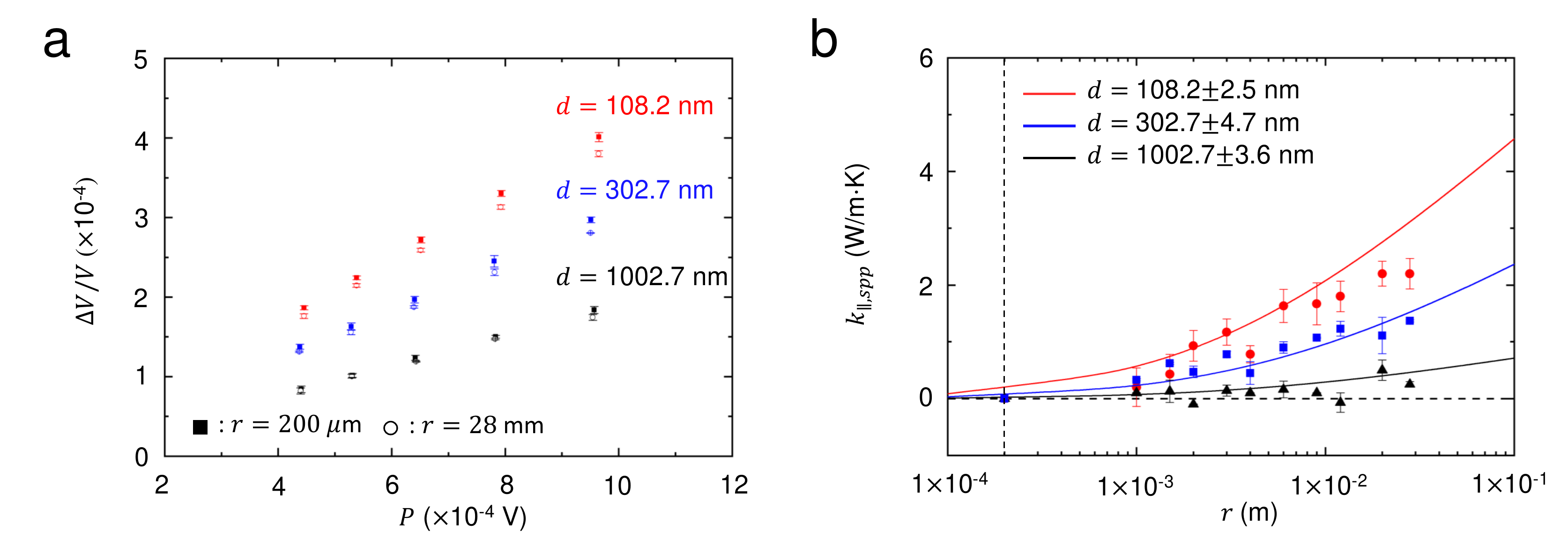}
\caption{SPP-enhanced thermal conductivity, $k_{\parallel,spp}$. (a) Measured probe reflectance response with respect to lock-in magnitude of pump laser photodetector response for the samples with different thickness and radius. Every plotted point was obtained by averaging the data over 5 seconds with time interval of 50 ms. The standard deviation of each point was within 1.5\%. (b) $k_{\parallel,spp}$ of the Ti film with respect to radius of the film. Sample with radius of 200 $\mu$m was used for calibration; that is, we set $k_{\parallel,spp}=0$ with $r=200$ $\mu$m for all samples. Every point is the averaged value of three repeated measurements.}\label{fig2} 
\end{figure}
%%%%%%%

Figure \ref{fig2}a shows the probe reflectance response ($\Delta V/V$) measured by the balanced photodetector for Ti films with different thicknesses and radii. The probe reflectance response of the Ti film with $r=28$ mm is noticeably lower than that of the Ti film with $r=200$ $\mu$m at a given thickness $d$. For example, $\Delta V/V$ for 108.2-nm-Ti film is 6.5\% lower when $r=28$ mm than $r=200$ $\mu$m. We believe that this difference in normalized probe reflectance with respect to the sample radius is caused by SPPs propagating along the Ti film for several millimeters.

Considering that the probe reflectance response is proportional to the surface temperature change $\Delta T$, and the absorbed heat $Q$ is proportional to the pump laser photodetector signal $P$, the proportionality constant $\gamma$ can be defined as \cite{braun2019steady}
\begin{equation} \label{Eq:1}
	 \left( \frac{\Delta V}{V P}\right)=\gamma \left( \frac{\Delta T(k)}{Q}\right).
\end{equation}
The thermal conductivity $k$ can be fitted to experimental data (i.e., $\Delta V/(VP)$) for a given $\gamma$ by calculating $\Delta T(k)/Q$ based on the 2-D heat diffusion model \cite{braun2017upper}. To determine the value of $\gamma$, one needs a calibration sample with known thermal conductivity (i.e., $k_{cal}$), i.e., 
\begin{equation} \label{Eq:2}
	 \gamma=\left(\frac{\Delta T(k_{cal})}{Q}\right)^{-1}_{cal}\left(\frac{\Delta V}{V P}\right)_{cal},
\end{equation}
where subscript `cal' implies calibration sample. In this work, the Ti film with the smallest radius (i.e., $r=200$ $\mu$m)  is used as the calibration sample because it has negligible SPP thermal conductivity (to be discussed in Fig. \ref{fig2}b). To obtain $\gamma$ from Eq. \\eqref{Eq:2}, the \textquoteleft\textit{intrinsic}\textquoteright\text{ } thermal conductivity of the Ti film and glass substrate must be known. For the Ti film, one needs to consider the anisotropic nature of  \textquoteleft\textit{intrinsic}\textquoteright\text{ } thermal conductivity due to its nanoscale thickness. To the in-plane direction, electronic contribution to the in-plane thermal conductivity ($k_{\parallel,e}$) was obtained by using the four-probe measurement and  Wiedemann\textunderscore Franz Law \cite{ashcroft1976solid}(Section 5 of Supplemental Material). Notice that $k_{\parallel,e}$ values in Table S3 are nearly independent of radius because the sample size is already several orders of magnitude larger than the mean free path of an electron ($\sim 10$ nm \cite{day1995correlation}). Therefore, the radius-dependence observed in Fig.\ \ref{fig2}a must be originated from other reasons than $k_{\parallel,e}$. The cross-plane thermal conductivity ($k_{\bot}$) of Ti film was separately measured by using the 3$\omega$ method (Fig.\ S5). The fitting procedures are described in Section 7 of Supplemental Material.

With the obtained $\gamma$ value, the in-plane thermal conductivity can be extracted from the measured $\Delta V/V$ values. The SSTR measures the in-plane thermal conductivity ($k_{\parallel}$) containing both contributions of electrons and SPPs, i.e., $k_{\parallel}=k_{\parallel, e} + k_{\parallel, spp}$. Therefore, a \textit {a priori} knowledge of $k_{\parallel, e}$ leads to $k_{\parallel, spp}$ from the SSTR measurements. Figure \ref{fig2}b plots the extracted $k_{\parallel, spp}$ from the SSTR measurements with respect to the sample radius for three selected thicknesses of 108.2 nm, 302.7 nm, and 1002.7 nm. The thicknesses of the samples were measured with a stylus profiler (Alpha-step 500, KLA TENCOR CORP), as shown in Table S5. First of all, $k_{\parallel, spp}$ in Fig. \ref{fig2}b exhibits strong radius-dependence except for the case of the thickest sample. For example, the 108.2-nm-thick Ti film can lead to considerable enhancement ($\sim 35\%$) in $k_{\parallel, spp}$ with $r=28$ mm, as compared to the case of the minimum $r$ of 200 $\mu$m. Such a high enhancement cannot be explained by the electronic contribution as electrical conductivity is nearly constant (Table S3). The theoretical prediction of $k_{\parallel, spp}$ by the Boltzmann transport equation is also plotted in Fig. \ref{fig2}b for comparison purposes. The predicted $k_{\parallel, spp}$ agrees well with SSTR measurements, indicating an SPP contribution to the in-plane thermal conductivity. Due to long propagation length of SPPs, $k_{\parallel, spp}$ has a strong radius dependence. Fig. \ref{fig2}a provides the first quantitative evidence for the ballistic nature of surface-polariton-enhanced in-plane thermal conductivity.

When Ti film thickness is 302.7 nm, $k_{\parallel, spp}$ is smaller than that with $d=108.2$ nm. For example, the 302.7-nm-thick Ti film has nearly half the $k_{\parallel, spp}$ of the 108.2-nm-thick Ti film at $r=28$ nm. If $d$ increases further to 1002.7 nm, there exists no noticeable $k_{\parallel, spp}$. Considering that the SPP thermal conductivity becomes significant only for nanoscale Ti films ($d \sim 100$ nm), the supported structure proposed in this work is clearly advantageous over the freestanding membrane structure for practical applications.

%%%%%%%%%%
\begin{figure}[t!]%
\centering
\includegraphics[width=1\textwidth]{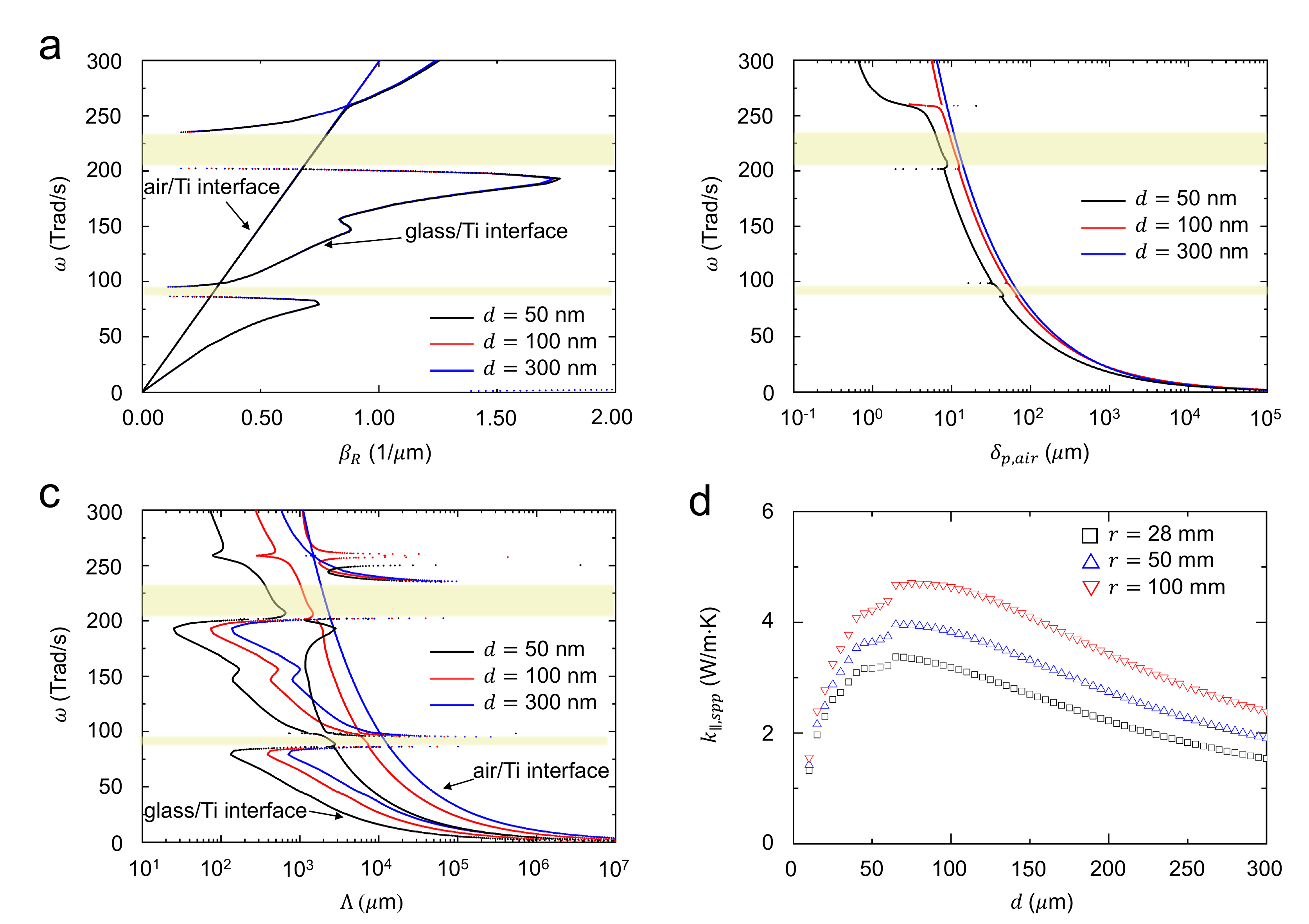}
\caption{SPP dispersion curves for Ti films on a glass substrate with film thickness $d=$ 50 nm, 100 nm, and 300 nm. (a) real part of in-plane wavevector ($\beta_R$), (b) skin depth, and (c) propagation length ($\Lambda$) of SPP at air/Ti interface penetrating into air with respect to film thickness. Shaded region indicates the frequency range where the real part of dielectric function of glass substrate is negative. In other words, SPPs originated from glass/Ti interface can not be excited in the shaded region. (d) Calculated $k_{\parallel, spp}$ with respect to the film thickness for different values of radius.}\label{fig3}
\end{figure}
%%%%%%%

Enhancement of the in-plane thermal conductivity of the Ti film can be explained by thermal transport via SPPs along a thin Ti film \cite{chen2005surface, ordonez2013anomalous}, which leads to 
\begin{equation} \label{Eq:3}
	 k_{\parallel, spp}=\frac{1}{4\pi d}\int_{0}^{\infty} k_\omega d\omega=\frac{1}{4\pi d}\int_{0}^{\infty} \hbar \omega \Lambda_{eff} \beta_R \frac{df_0}{dT} d\omega
\end{equation}
where $\hbar$ is Planck constant divided by 2$\pi$, $\omega$ is the angular frequency, $\beta_R$ is the real part of the in-plane wavevector (i.e., $\beta=\beta_R+i\beta_I$), and $f_0$ is the Bose-Einstein distribution function. Because the propagation length of SPPs is defined by the imaginary part of the in-plane wavevector, i.e., $\Lambda=1/(2\beta_I)$ \cite{chen2005surface, ordonez2013anomalous}, which in turn leads to $\Lambda_{eff}$, one need to solve the dispersion relation of SPPs supported by a thin Ti film deposited on a glass substrate \cite{ordonez2014thermal, lim2019thermal}:
\begin{equation} \label{Eq:4}
	 \tanh (p_{\text{Ti}}d)=-\frac{p_{\text{Ti}}\varepsilon_{\text{Ti}}(p_a\varepsilon_s+p_s\varepsilon_a)}{p_{\text{Ti}}^2\varepsilon_a\varepsilon_s+p_ap_s\varepsilon_{\text{Ti}}^2}
\end{equation}
where `$a$' and `$s$' implies the air and substrate, respectively, and $p_i$ is the cross-plane wavevector of the $i^{th}$ medium (i.e., $p_i^2=\beta^2-\varepsilon_ik_0^2$ with $\varepsilon_i$ being the dielectric function of the medium $i$). The dielectric function of glass (i.e., amorphous SiO$_2$) is taken from tabulated data \cite{palik1998handbook}. For simple calculation, the real part of the dielectric function was used for calculation of SPP dispersion \cite{raether1988surface}.

The real part of in-plane wavevector $\beta_R$ of Ti films deposited on glass substrates with $d=50$ nm, 100 nm, and 300 nm is plotted in Fig.\ \ref{fig3}a. The dispersion curves are drawn for a frequency range from 0 to 300 Trad/s because the calculated spectral thermal conductivity of the Ti film on the glass substrate (i.e., $k_\omega=1/(4\pi d)\hbar\omega\Lambda_{eff}\beta_R(df_0/dT)$) has considerable value at frequency lower than 300 Trad/s when maximum propagation length, $L$ is set to be 28 mm (see Fig.\ S8). In Fig.\ \ref{fig3}a, it is clearly seen that there exist two branches of SPP dispersion, which are originated from the SPPs at two interfaces of Ti film. For metal film with asymmetric surrounding dielectrics, symmetric mode which has longer propagation length is supported at the interface between metal and dielectric with lower refractive index (i.e., lower $\beta_R$ at a given frequency)  \cite{burke1986surface}. In other words, the symmetric mode of SPPs is supported at the air/Ti interface and has longer propagation length than that supported at the glass/Ti interface (see Fig.\ \ref{fig3}c). Because much longer propagation can be obtained for SPPs from air/Ti interface, following discussion will be mainly focused on the SPPs at the air/Ti interface. 

At the interface between a semi-infinite air and a semi-infinite Ti, a SPP dispersion follows the light line of air (i.e., photon-like behavior) in the sufficiently low frequency region before reaching an asymptote at high frequency ($\sim \omega_p/\sqrt{2}$ with $\omega_p$ being the plasma frequency of Ti) \cite{raether1988surface}. Although all of the $\beta_R$ dispersion curves at the air/Ti interface look similar and coincide with the air light line regardless of the thickness of the Ti film in Fig.\ \ref{fig3}a, there exists a slight difference near the air light line for different thicknesses of Ti films (Fig. S9). As $d$ decreases, the dispersion deviates from the air light line due to the coupling of the SPPs supported at the air/Ti interface and those at the glass/Ti interface. Importantly, the deviation of $\beta_R$ from the light line leads to the confinement of SPPs at the interface. Fig.\ \ref{fig3}b depicts the skin depth of SPPs from the air/Ti interface (i.e., the penetration depth into air) with respect to film thickness, where $\delta_{p,air}=1/[2\text{Re}(p_a)]$. As the separation between $\beta_R$ and the light line becomes smaller, the field can penetrate deeper into surrounding air, resulting in a longer propagation length, as in Fig.\ \ref{fig3}c. Thus, SPPs with a long propagation length exist for thick Ti films where the SPPs at the air/Ti interface are located close to the light line. Although not shown here, $\Lambda$ with $d=1000$ nm is essentially overlapped with that with $d=300$ nm, suggesting that 300-nm-thick Ti film is already too thick for coupling of SPPs at both interfaces.

For a freestanding SiO$_2$ membrane, surface waves propagate longer when the membrane is thinner \cite{tranchant2019two}. Such discrepancies result from the different optical properties of a lossy metal and a polar dielectric. For a SiO$_2$ membrane, the Zenneck surface modes ($\epsilon_r > 0$ and $\epsilon_i > 0$, where $\epsilon_r$ and $\epsilon_i$ represent real and imaginary parts of the permittivity of the thin film) dominate thermal conductivity by surface waves, while metal thin films can support SPPs (i.e., $\epsilon_r < -\epsilon_{air}$) in a broad frequency range. Thus, the energy loss of surface waves decreases as film thickness decreases for SiO$_2$ membrane, leading to a longer propagation length. Also, for a metal film surrounded by same dielectrics, a symmetric SPP mode approaches the light line as the thickness of the metal film decreases and the coupling of SPPs at each metal/dielectric interface becomes significant \cite{burke1986surface, yang1991long}, resulting in longer propagation length of SPPs.

Contrarily, for a thin metal film on a dielectric substrate, SPPs can propagate longer along the interface when the film thickness is greater than the penetration depth of the metal film, so SPPs at the air/Ti interface can approach the air light line. In fact, when the analytic solution for the SPP dispersion at the single interface of semi-infinite metal/dielectric \cite{burke1986surface}is applied, the resulting dispersion is almost equal to that of a metal film with $d=300$ nm (Section 11 of Supplemental Material), which implies that the SPPs of each interface are nearly decoupled and shows the characteristics of a semi-infinite metal/dielectric interface.

At low frequencies below 100 Trad/s, the SPP propagation lengths can be longer than 1 cm, causing the radius-dependence of $k_{\parallel, spp}$ in Fig. \ref{fig2}b. Considering that the effective propagation length of SPP is given by $\Lambda_{eff}^{-1}=\Lambda^{-1}+{r}^{-1}$, the radius of samples mostly limits low-frequency SPPs under 100 Trad/s if $r<1$ cm. Equation \eqref{Eq:3} also suggests that $1/(4\pi d)$ also determines the SPP thermal conductivity. This is the reason why the 302.7-nm-thick Ti film has lower $k_{\parallel, spp}$ than the 108.2-nm-thick Ti film even though it has longer $\Lambda$. In other words, the contribution of thermal conduction via SPPs can be substantial in thin films, although the SPPs can propagate longer for thicker films, implying the existence of an optimum thickness for SPP-enhanced thermal conductivity.

Figure \ref{fig3}d shows the calculated $k_{\parallel, spp}$ with respect to Ti film thickness when $r=28$ mm (i.e., maximum sample size), 50 mm, and 100 mm. Because of the interplay between the factor of $1/(4\pi d)$ and $\beta_R \Lambda_{eff}$ in Eq.\ \eqref{Eq:3}, the optimal thickness $d$ exists at a given radius $r$. For the considered Ti film deposited on a glass substrate, the optimal thickness is approximately 70 nm. As $d$ decreases below 70 nm, the reduction of SPP propagation length (due to coupling of SPPs at both interfaces) results in decreased $k_{\parallel, spp}$. On the other hand, as $d$ increases beyond 70 nm, the factor of $1/(4\pi d)$ will eventually lead to decreased $k_{\parallel, spp}$. Because the radiation penetration depth of Ti is approximately 40 nm in the infrared spectral region ($<200$ Trad/s), the coupling of SPPs may occur up to $d \sim 120$ nm (i.e., three times the radiation penetration depth). 

In conclusion, the SSTR measurements for the supported Ti film on glass substrate clearly demonstrate the thickness- and size-dependence of the surface-polariton-enhanced in-plane thermal conductivity. For the 108.2-nm-thick Ti film, the corresponding $k_{\parallel, spp}$ is nearly 40\% of its intrinsic value (i.e., $k_{\parallel, e}$) when $r=28$ mm. Due to broadband SPPs supported at an air/lossy metal interface, it is found that there exists an optimal thickness for the maximum $k_{\parallel, spp}$. This study provides a practical approach for exploiting surface polariton to enhance in-plane thermal conductivity by introducing a supported metal thin film configuration, where the enhancement of thermal conductivity is almost impossible for polar-dielectric counterparts. 

\begin{acknowledgments}
\section*{Supplemental Material}
See Supplemental Material at [URL will be inserted by publisher] for (1) Validation of experimental setup, (2) Sensitivity analysis of SSTR, (3) Summary of thermal properties used in 2-D heat diffusion model, (4) Optical property of Ti film, (5) Electron contribution of in-plane thermal conductivity of Ti film ($k_{\parallel, e}$) measured by 4-probe measurement, (6) Measurement of cross-plane thermal conductivity of Ti film by using 3$\omega$ method, (7) Fitting procedure and uncertainty of $k_{\parallel, spp}$ obtained by SSTR method, (8) Measurement of Ti film thickness with stylus profiler, (9) Spectral thermal conductivity of SPP, (10) Dispersion curve near the light line of air, (11) Analytic solution of SPP, (12) Beam diameter of probe and pump laser, and (13) Thermal conductivity of glass substrate. All files related to a published paper are stored as a single deposit and assigned a Supplemental Material URL. This URL appears in the article\textquoteright s reference list.

\section*{Funding Sources}
This research is supported by the Basic Science Research Program (NRF-2020R1A4 A4078930) through the National Research Foundation of Korea (NRF) funded by Ministry of Science and ICT.

\end{acknowledgments}


\bibliography{Kim_aps.bib}

\end{document}

\setstretch{1.5}
\section*{Supplemental Material:\\
Boosting thermal conductivity by surface plasmon polaritons propagating along a thin Ti film}

\textbf{Authors}: Dong-min Kim$^{1,2}$, Sinwoo Choi$^{1,2}$, Jungwan Cho$^{3}$, Mikyung Lim$^{4*}$, Bong Jae Lee$^{1,2*}$\\

\textit{1. Department of Mechanical Engineering, Korea Advanced Institute of Science and Technology, Daejeon 34141, South Korea\\
2. Center for Extreme Thermal Physics and Manufacturing, Korea Advanced Institute of Science and Technology, Daejeon 34141, South Korea\\
3. School of Mechanical Engineering, Sungkyunkwan University, Suwon 16419, South Korea\\
4. Nano-Convergence Manufacturing Systems Research Division, Korea Institute of Machinery and Materials, Daejeon 34103, South Korea\\
}\\
*Corresponding authors\\
*e-mail: mklim@kimm.re.kr (Mikyung Lim), bongjae.lee@kaist.ac.kr (Bong Jae Lee)\\

\bigskip
\bigskip
\bigskip
\underline{\textbf{Table of Contents}}\\
1. Validation of experimental setup\\
2. Sensitivity analysis of SSTR\\
3. Summary of thermal properties used in 2-D heat diffusion model\\
4. Optical property of Ti film\\
5. Electron contribution of in-plane thermal conductivity of Ti film ($k_{\parallel,e}$) measured by 4-probe measurement\\
6. Measurement of cross-plane thermal conductivity of Ti film by using 3$\omega$ method\\
7. Fitting procedure and uncertainty of $k_{\parallel,spp}$ obtained by SSTR method\\
8. Measurement of Ti film thickness with stylus profiler\\
9. Spectral thermal conductivity of SPP\\
10. Dispersion curve near the light line of air\\
11. Analytic solution of SPP\\
12. Beam diamter of probe and pump laser\\
13. Thermal conductivity of glass substrate\\

\setstretch{1.5}

\clearpage

\section*{1. Validation of experimental setup}

A custom-built steady-state thermoreflectance method \cite{braun2019steady} is used to measure the Ti film in-plane thermal conductivity. The SSTR is a variation of the frequency-domain thermoreflectance method (FDTR), which heats the sample to make it nearly at steady-state by substantially reducing the modulation frequency of the pump laser compared to the FDTR. The probe reflectance response ($\Delta V/V$) with respect to the pump laser modulation ($P$) is measured to obtain the in-plane thermal conductivity of Ti film. Here, $P$ is measured with a lock-in amplifier (SR830, Stanford Research) by monitoring a partly-picked-pump beam using a 90:10 beam splitter. The pump laser (Coherent OBIS 532, $\lambda=532$ nm) is modulated with a frequency of 500 Hz, which is expected to heat the Ti film until it rises over 90\% of its steady-state value (Fig.\ \ref{FigS8}a). The reflectance of the constant-wave probe laser (Coherent CUBIC 660, $\lambda=660$ nm) oscillates with the modulation frequency of the pump laser because the normalized reflectance of the metal transducer is proportional to its temperature ($T$). The waveform of the pump laser is a fully on/off square wave to mimic the constant heat flux boundary condition while deriving the steady-state temperature of the samples. Given that the change of the reflectance with respect to temperature is less than 1\% for a temperature change of 10 K, the DC signal of the probe ($V$) is eliminated to enhance the measurement sensitivity of $\Delta V$ by subtracting the reference signal from $V$ using a balanced photodetector (PDB415A, Thorlabs). A reference beam is extracted from the probe beam by splitting it through a polarized beam splitter. Two split probe beams are separately monitored by two photodiodes on a balanced photodetector. On the other hand, $V$ is measured by reading the DC signal of the channel where the probe beam is received. The beam size of the probe and pump laser after an objective lens (40x magnification with NA$=0.75$; UPLFLN 40X, Olympus) is about 5 $\mu$m, which was measured using the knife edge method (see Section 12).\\

\begin{figure}[t]
\centering\includegraphics[width=17cm]{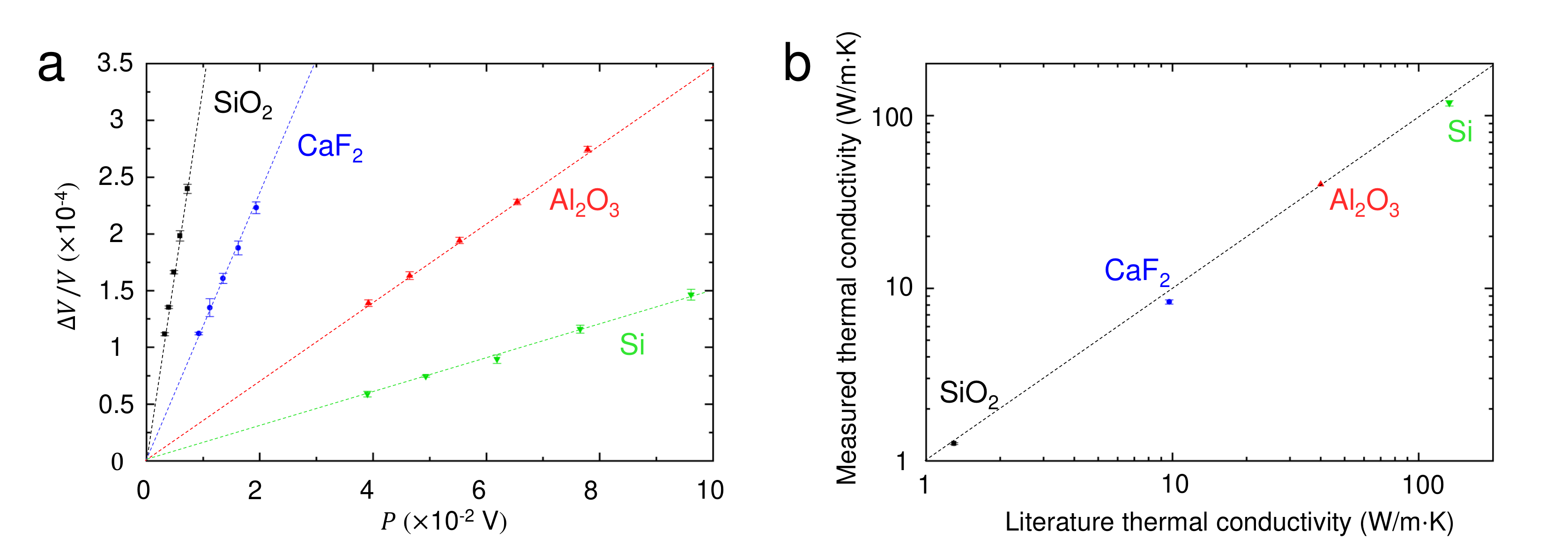}
\caption{\label{FigS1}Figure S1: (a) Probe reflectivity response with respect to lock-in magnitude of pump power for four reference materials. Every point was obtained by averaging three measurements. (b) Comparison between measured thermal conductivity and reported value for four reference materials.}
\end{figure}

 An experimental setup is validated by estimating the thermal conductivities of known materials. Here, four materials (SiO$_2$, CaF$_2$, Al$_2$O$_3$, and Si) are selected as reference materials. Among them, Al$_2$O$_3$ is used as a calibration sample for deriving $\gamma$, given that the sensitivity in measuring thermal conductivity of Al$_2$O$_3$ is higher than that of the others. An 80-nm-thick Al film is deposited on the reference sample and used as a metal transducer, and its themal conductivity is assumed to be 100 W/m$\cdot$K \cite{braun2019steady}. The modulation frequency of the pump laser is set as 200 Hz, where the detailed procedure for setting the pump laser frequency is similar to that described in Supplementary Note 12. The measured probe reflectance response ($\Delta V/V$) with respect to the pump laser photodetector signal ($P$) is plotted in Fig. S\ref{FigS1}a. The value of $\gamma$ is first estimated from the measured $\frac{\Delta V}{V P}$  of the  Al$_2$O$_3$ sample with known conductivity of $k=35$ W/m$\cdot$K \cite{braun2019steady}. Then, with $\frac{\Delta V}{V P}$ of other materials, $\frac{\Delta T(k)}{Q}$ can be calculated with the $\gamma$ of Al$_2$O$_3$. Thermal conductivity can be derived by letting it as a fitting parameter to theoretical values obtained from the 2-D heat diffusion model (Table S\ref{Tab:2}). As shown in Fig. S\ref{FigS1}, $\frac{\Delta V}{V P}$ decreases as the thermal conductivity of the substrate increases. It implies that more power is required to heat the Al surface when the substrate has higher thermal conductivity. All measured values show good agreement from their reported values, which confirms the reliability of our custom-built experimental setup.
 \newpage
%%%%%%%%%%%%%%%%%%%%
%%%%%%%%%%%%%%%%%%%%%%%

%%%%%%%%%%
\begin{table}[t]
	\caption{Table S1. Measured thermal conductivity of reference materials.}
	\label{Tab:2}
	\small
	\begin{center}
		\begin{tabular} {c| c |  c |  c | c |c   } 
		\hline\hline
		Material & $\Delta T_{ss}$ & $\gamma$ & $\frac{\Delta V}{V\times P}$ & $k_{exp}$ & Ref. \cite{braun2019steady} \\
	             & (K) &  & ($\text{V}^{-1}$) & (W/m$\cdot$K) & (W/m$\cdot$K)\\
		\hline
     	SiO$_2$ & $<20$ & 3.3$\times10^{-6}$ & 0.033 & 1.2$\pm$0.01  & 1.3 \\
     	CaF$_2$ & $<20$ & 3.3$\times10^{-6}$ & 0.011 & 7.65$\pm$0.19  & 9.7  \\    
     	Al$_2$O$_3$ (calib.)& $<20$ & 3.3$\times10^{-6}$ & 0.0035 & - & 35  \\    
     	Si & $<20$ & 3.3$\times10^{-6}$ & 0.0015 & 100$\pm$3.2 & 133 \\                 
   		\hline\hline    
		\end{tabular}
	\end{center}
\end{table} 
%%%%

\clearpage

\section*{2. Sensitivity analysis of SSTR}

\begin{figure}[h]
\centering\includegraphics[width=17cm]{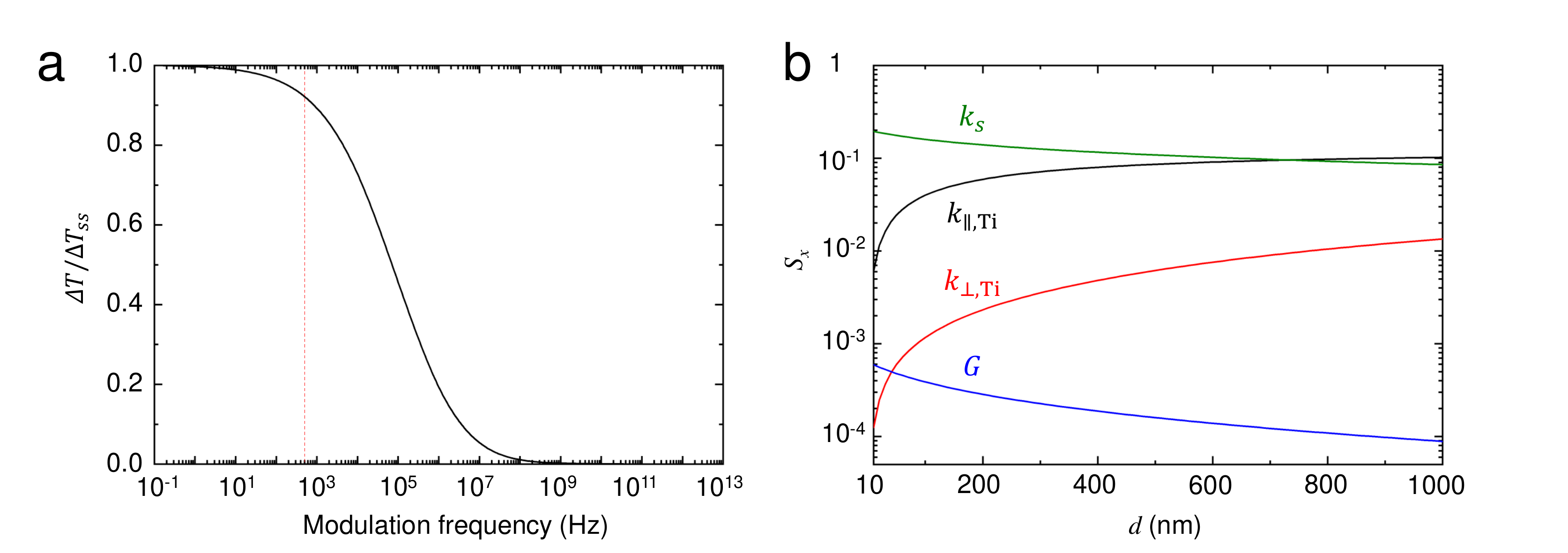}
\caption{\label{FigS8}Figure S2: (a) Temperature rise normalized with steady-state temperature with respect to modulation frequency. (b) Sensitivity of input parameters of 2-D heat diffusion model with respect to Ti film thickness. }
\end{figure}

The modulation frequency of the pump laser for thermal conductivity measurement of Ti film was determined by the 2-D heat diffusion model \cite{braun2017upper} as shown in Fig. S\ref{FigS8}a. The temperature rise of the sample with respect to modulation frequency was calculated. Here, thermal properties in Table S\ref{Tab:1} were used and $k_{\parallel,e}$ was assumed to be 10 W/m$\cdot$K. As a result, we set the pump laser modulation frequency as 500 Hz because the temperature of the sample surface rises over 90\% of its steady-state temperature at the corresponding frequency. Figure S\ref{FigS8}b shows the sensitivity ($S_x$) of thermal properties with respect to film thickness $d$, where $S_x$ is defined as  
\begin{equation} \label{Eq:5}
	S_x=\frac{|\Delta T_{1.1x}(d)-\Delta T_{0.9x}(d)|}{\Delta T_x(d)}
\end{equation}
where $\Delta T_x$ is the temperature rise derived from the input parameter $x$. As shown in the figure, substrate thermal conductivity ($k_s$, thermal conductivity of the glass substrate) and in-plane thermal conductivity of Ti film ($k_{\parallel,Ti}$) have relatively high sensitivity, while cross-plane thermal conductivity of Ti film ($k_{\perp,Ti}$) and boundary conductance ($G$) show low sensitivity. This result implies that $k_{\perp,Ti}$ and $G$ do not have significant effects on deriving the $k_{\parallel,Ti}$ from the experimental data and the 2-D heat diffusion model. Thus, it gives justification to approximate $k_{\perp,Ti}$ as measured data of Ti film with $d=100$ nm from Section 6, and $G$ as 200 MW/m$^{2}\cdot$K \cite{braun2019steady} in the 2-D heat diffusion model. 

%%%%%%%%%%%%%%%%%%%%
%%%%%%%%%%%%%%%%%%%%%%%

\newpage

\section*{3. Summary of thermal properties used in 2-D heat diffusion model}

We employed the 2-D heat diffusion model in cylindrical coordinates with a radial symmetry and a Gaussian profile of surface heating source \cite{braun2017upper}: 
\begin{equation} \label{Eq:1}
  k_{\parallel} \left \{ \frac{1}{r} \frac{\partial T(z,r,t)}{\partial r}+\frac{\partial^2 T(z,r,t)}{\partial r^2} \right \}+k_{\perp}\frac{\partial^2T(z,r,t)}{\partial z^2}=C_v\frac{\partial T(z,r,t)}{\partial t},
\end{equation}
where $r$ denotes radius (in-plane) of coordinate, $z$ is cross-plane depth orthogonal to $r$, $C_v$ refers the volumetric heat capacity, and $t$ is the time. The boundary condition at $z=0$ is given by surface heating condition, i.e., 
\begin{equation} \label{Eq:2}
  \frac{\partial T(z,r,t)}{\partial z}\bigg |_{z=0}=- \frac{1}{k_{\perp}} Q(r)
\end{equation}
where $Q(r)$ indicates the heat flux due to CW laser given by
\begin{equation} \label{Eq:3}
  Q(r)=\frac{2}{\pi r_0^2}\exp\left(-\frac{2r^2}{r_0^2}\right)\alpha A
\end{equation}
In the above equation, $r_0$ refers to the $1/e^2$ radius of the pump laser, $\alpha$ is the absorptivity and $A$ is the laser power. To derive the surface temperature $T(0,r,t)$ from Eq.\ \eqref{Eq:3}, the semi-infinite boundary condition is applied at the bottom side of the structure, i.e.,  
\begin{equation} \label{Eq:4}
  \frac{\partial T(z,r,t)}{\partial z}\bigg |_{z \rightarrow \infty}= 0
\end{equation}
Note that 500-$\mu$m-thick glass wafer used for the measurements can be assumed to be semi-infinite, because its thermal penetration depth is less than 10 $\mu$m when the beam diameter is 5 $\mu$m \cite{braun2017upper}. \\

Thermal properties used in solving the 2-D heat diffusion model are summarized in Table S\ref{Tab:1}. The film thickness of the Ti film measured with a stylus profiler (Alpha-Step 500, KLA TENCOR CORP.) is also listed. The electron contribution of the Ti film in-plane thermal conductivity ($k_{\parallel,e}$) is derived from the in-plane electrical conductivity of the Ti film measured with the 4-probe method (4200-SCS, Keithley) and the Lorentz number of Ti (2.75$\times10^{-8}$ W$\cdot\Omega$/K$^2$ \cite{ashcroft1976solid}). The cross-plane thermal conductivity of the Ti film ($k_{\perp}$) is estimated with a 3$\omega$ method \cite{borca2001data}. Thermal conductivity of the glass substrate is calculated from thermal diffusivity measured with a laser flash analysis (Netzsch LFA 457) and heat capacity obtained with a differential scanning thermometry (Netzsch DSC 204 F1 Phoenix). Values for volumetric heat capacitance ($C_v$) of the glass substrate and the Ti film are adopted from \cite{braun2019steady} and \cite{matsui2011analysis}, respectively. Note that due to the low sensitivity of boundary conductance $G$ in solving the 2-D heat diffusion model (see Fig.\ S\ref{FigS8}b), the value is assumed to be 200 MW/m$^{2}\cdot$K as in \cite{braun2019steady}. Detailed procedures for obtaining the thermal properties of the materials are described in the corresponding supplementary note. \\

\clearpage

%%%%%%%%%%
\begin{table}[h]
	\caption{Table S2. Summary of the thermal properties used in solving 2-D heat diffusion model.}
	\label{Tab:1}
	\small
	\begin{center}
		\begin{tabular} {c | c | c | c | c | c | c | c } 
		\hline\hline
		 & Film thickness & $k_{\parallel,e}$ \footnotemark[1]& $k_{\perp,e}$\footnotemark[2] & $k_{\text{SiO$_2$}}$\footnotemark[3] &$C_{v,\text{SiO$_2$}}$ \cite{braun2019steady} & $C_{v,\text{Ti}}$ \cite{matsui2011analysis} & $G$ \cite{braun2019steady} \\
	             & (nm) & (W/m$\cdot$K) & (W/m$\cdot$K)  & (W/m$\cdot$K)  & (MJ/m$^{3}\cdot$K) & (MJ/m$^{3}\cdot$K) &  (MJ/m$^{3}\cdot$K) \\
		\hline
     	Sample 1 & $108.2\pm2.5$ & $7.14\pm0.16$ & & & & & \\
     	Sample 2 & $302.7\pm4.7$ & $10.6\pm0.34$ & $5.88\pm0.28$ & $1.35\pm0.04$ & 1.66 & 2.35 & 200 \\  
     	Sample 3   & $1002.7\pm3.6$ & $8.46\pm0.64$ & & & & & \\                                      
   		\hline\hline    
		\end{tabular}
	\end{center}
\end{table} 
%%%%

\footnotetext[1]{See section 5}
\footnotetext[2]{See section 6}
\footnotetext[3]{See section 13}

\clearpage

\section*{4. Optical property of Ti film}

\begin{figure}[h]
\centering\includegraphics[width=17cm]{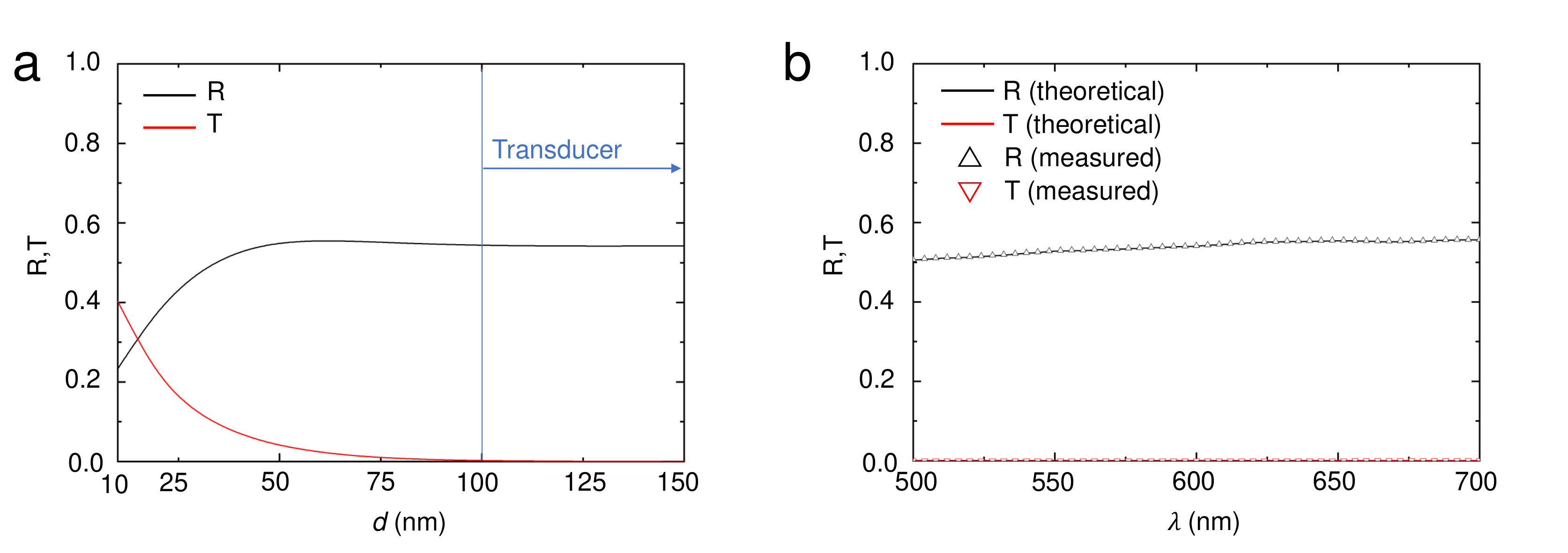}
\caption{\label{FigS2} Figure S3: (a) Calculated reflectance $R$ and transmittance $T$ of Ti film with respect to film thickness. (b) Calculated and measured spectral reflectance and transmittance of Ti film in wavelength range lies in 500 nm to 700 nm.}
\end{figure}

When measuring the thermal conductivity of a Ti film, the Ti film itself is used as a metal transducer at the same time. Because a semi-transparent film can induce multiple reflections inside the film, reflectance varies depending on the film thickness. Thus, in order to use a metal film as a transducer for the thermoreflectance method, a considerate selection of the metal film thickness that makes it opaque is necessary for reliable measurement regardless of film thickness uncertainty. The spectral reflectance and transmittance of a Ti film on a glass substrate are calculated with respect to the film thickness using Airy's multilayer formulas \cite{zhang2007nano} where the frequency-dependent dielectric function of Ti film is obtained from the Drude model \cite{ashcroft1976solid,palik1998handbook}. As shown in Fig. S\ref{FigS2}, Ti film becomes opaque when the film thickness is greater than 100 nm. Thus, we set $d=100$ nm as the minimum Ti film thickness for the SSTR measurement. To ensure the applicability of Drude-model-based optical properties, the spectral reflectance and transmittance of the 100-nm thick Ti film deposited on a glass substrate are measured with a UV-VIS spectrometer (UV-3600 Plus, Shimadzu) combined with an integrating sphere (ISR-603, Shimadzu) at wavelengths from 500 nm to 700 nm, which includes the pumping wavelength (i.e., $\lambda=532$ nm) and the probing wavelength (i.e., $\lambda=660$ nm). In Fig. S\ref{FigS2}b, it can be readily noted that the measured spectrum agreed excellently with the theoretical prediction based on the Drude model. Here, the Drude model parameters of the Ti film are $\omega_\tau=372$ cm$^{-1}$ and $\omega_p=25,000$ cm$^{-1}$ \cite{ordal1988optical}.

%%%%%%%%%%%%%%%%%%%%
%%%%%%%%%%%%%%%%%%%%%%%

\newpage

\section*{5. Electron contribution of in-plane thermal conductivity of Ti film ($k_{\parallel,e}$) measured by 4-probe measurement}

\begin{figure}[h]
\centering\includegraphics[width=17cm]{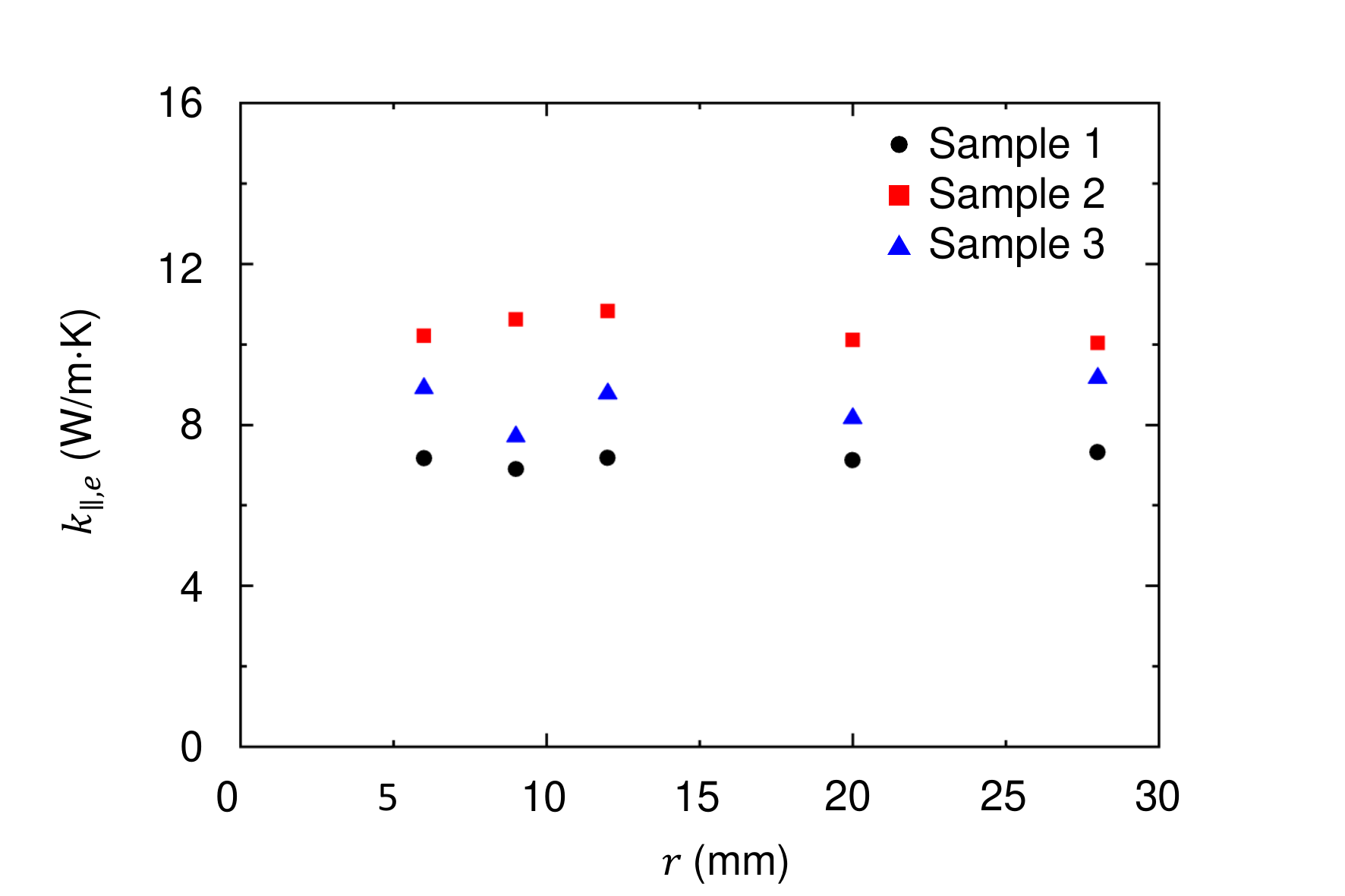}
\caption{\label{FigS3}Figure S4: Electron contribution of in-plane thermal conductivity of Ti films ($k_{\parallel,e}$) with respect to film radius $r$.}
\end{figure}

For a given Ti-film thickness ($d=108.2$, $302.7$, and $1002.7$ nm), the Ti film was patterned by the lift-off process with circular shapes of a total of 11 different radius ($r$) values ranging from 200 $\mu$m to 28 mm. The electron contribution of the in-plane thermal conductivity of Ti films ($k_{\parallel,e}$) is derived from its sheet resistance measured by the 4-probe method (4200-SCS, Keithley). Metals, including Ti, follow the Wiedemann-Franz law, which states that thermal conductivity and electrical conductivity are proportional as
\begin{equation} \label{Eq:5}
	\frac{k_{\parallel,e}}{\sigma}=L_{\text{Ti}}T.
\end{equation}
In the above equation, $L_{\text{Ti}}$ is the Lorentz number of Ti which is known as 2.75$\times 10^{-8}$ W$\cdot\Omega$/K$^2$ \cite{ashcroft1976solid}, $T$ is temperature, and $\sigma$ represent the electrical conductivity of the metal, which can be obtained from the measured sheet resistance $R_s$ as follows:
\begin{equation} \label{Eq:6}
	\rho=\frac{1}{\sigma}=R_s d
\end{equation}
where $\rho$ refers to the electrical resistivity and $d$ denotes film thickness. The $k_{\parallel,e}$ of the film derived from Eq. \eqref{Eq:5} is shown in Table S\ref{Tab:3} and Fig. S\ref{FigS3}. Due to the minimum spacing limit of the probes, measurements are performed on films with a radius of greater than 6 mm. Since the three samples were deposited in different batches, the electrical conductivity of each sample may differ slightly depending on the deposition environment as shown in Fig.\ S\ref{FigS3}. In the case of Sample 1 ($d=108.2$ nm), the standard deviation of $k_{\parallel,e}$ is found to be 0.15 W/m$\cdot$K (2\% of average value), while the maximum thermal conductivity enhancement by surface plasmon polaritons (SPPs) is $\sim$2.5 W/m$\cdot$K, which implies a marginal effect of uncertainty of $k_{\parallel,e}$ on derivation of $k_{\parallel,spp}$. $k_{\parallel,e}$ of Sample 2 ($d=302.7$ nm) and Sample 3 ($d=1002.7$ nm) shows a standard deviation of 3.3\% and 7\%, respectively. In the case of Sample 2, the standard deviation is also relatively small compared to the $k_{\parallel,spp}$, which is enhanced up to 15\% of its bulk value. In addition, because each sample consists of two wafers, the standard deviation of Sample 3 can be reduced to around 2\% when considering $k_{\parallel,e}$ variation within one wafer.\\

%%%%%%%%%%
\begin{table}[!h]
	\caption{Table S3. $k_{\parallel,e}$ of Ti film for $r \geq 6$ mm.}
	\label{Tab:3}
	\small
	\begin{center}
	\begin{tabular}{c| c |  c |  c}
	\hline
	\hline
	& \multicolumn{3}{c}{$k_{\parallel,e}$ (W/m$\cdot$K)}             \\
	 \hline
$r$ (mm)  & Sample 1 & Sample 2 & Sample 3 \\
\hline
6       & 7.17     & 10.22    & 8.91     \\
9       & 6.9      & 10.62    & 7.71     \\
12      & 7.18     & 10.83    & 8.78     \\
20      & 7.12     & 10.11    & 8.17     \\
28      & 7.32     & 10.04    & 9.17     \\
\hline
average & 7.14    & 10.36   & 8.55    \\
stdev.  & 0.15 & 0.34 & 0.60\\
\hline
    \hline
\end{tabular}
\end{center}
\end{table}
%%%%%%%%%%%%

%%%%%%%%%%%%
%%%%%%%%%%%%%

\clearpage

\section*{6. Measurement of cross-plane thermal conductivity of Ti film by using 3$\omega$ method}

\begin{figure}[h]
\centering\includegraphics[width=12cm]{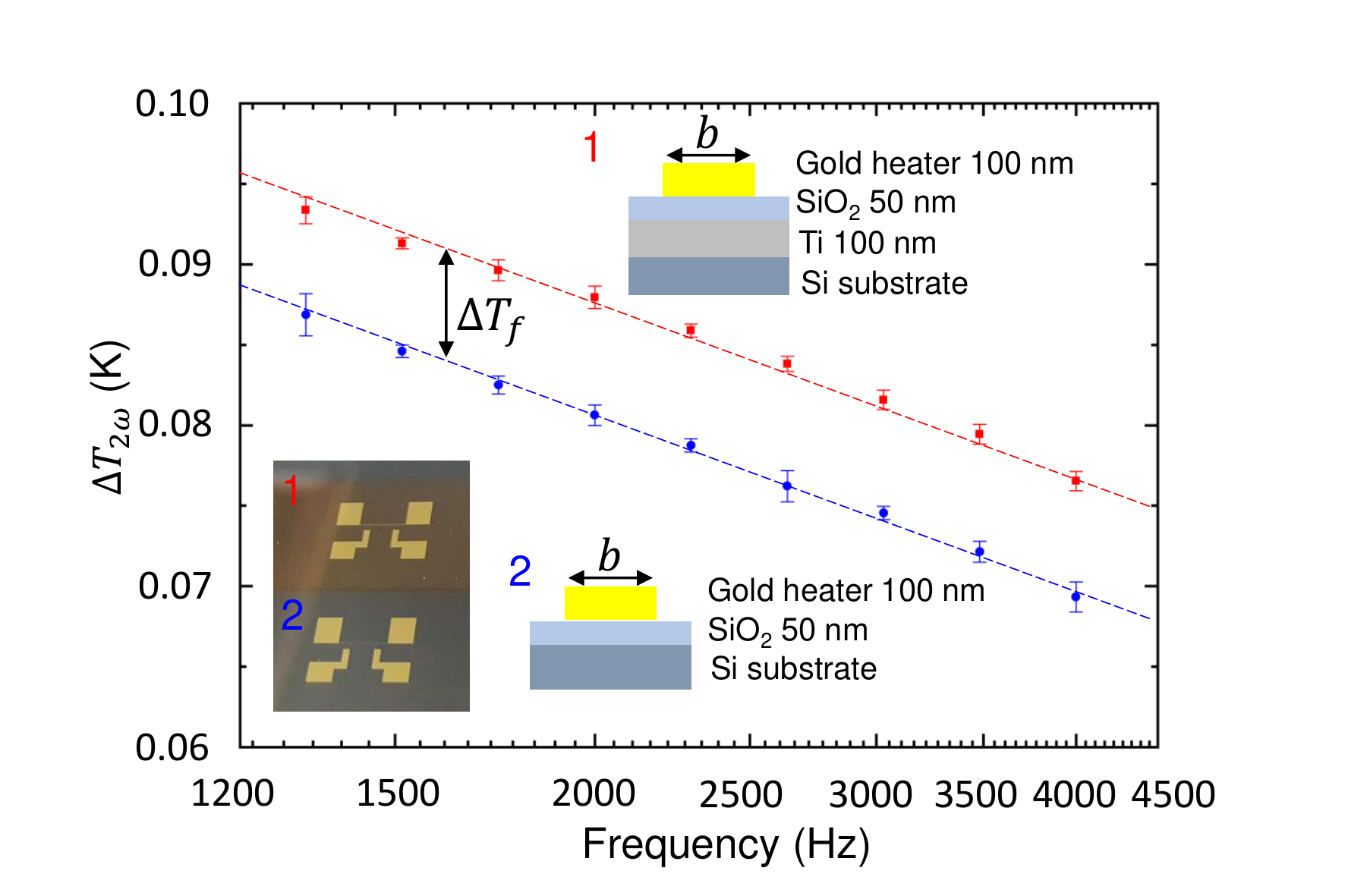}
\caption{\label{FigS4}Figure S5: Temperature oscillation $\Delta T_{2\omega}$ of heater with respect to heating frequency of measurement sample (Sample 1) and reference sample (Sample 2).}
\end{figure}

Electron contribution of the cross-plane thermal conductivity of Ti film can be measured with the 3$\omega$ method. Here, the differential method, which is a variation of the 3$\omega$ method, is introduced to derive the cross-plane thermal conductivity of Ti film \cite{borca2001data}. In this method, the cross-plane thermal conductivity of thin film can be obtained by measuring the temperature difference ($\Delta T_f$) between samples with and without Ti film (i.e., Sample 1 and Sample 2 in Fig. S\ref{FigS4}). For the differential method, one-dimensional (1-D) heat conduction should be assumed for Ti film to obtain the film thermal conductivity from $\Delta T_f$. By using a substrate with a high thermal conductivity (Si, $k=133$ W/m$\cdot$K \cite{braun2019steady}) and a heater with a wider width than the Ti film thickness ($b=30$ $\mu$m), Ti film well satisfies the 1-D heat conduction condition \cite{borca2001data}. Temperature oscillation ($\Delta T_{2\omega}$) of two samples with respect to heating frequency is shown in Fig.S\ref{FigS4}. The slope of $\Delta T_{2\omega}$ according to frequency is determined by the thermal conductivity of the substrate. As shown in Fig.S\ref{FigS4}, it can be seen that the slope of the two samples is nearly consistent, and there is only a temperature offset induced by the Ti film, meaning that 1-D heat resistance exists across the Ti film. As a result, the $k_{\perp,e}$ was obtained as $5.88\pm0.28$ W/m$\cdot$K by using Fourier's law
\begin{equation} \label{Eq:7}
	Q=k_{\perp,e}bl\frac{\Delta T_f}{d}
\end{equation}
where Q is applied power to the line heater, $d$ is film thickness, and $l$ is length of the line heater (i.e., $l=$2 mm).
%%%%%%%%%%%%%%%%%%
%%%%%%%%%%%%%%%%%%%%
\clearpage

\section*{7. Fitting procedure and uncertainty of $k_{\parallel,spp}$ obtained by SSTR method }

\begin{figure}[h]
\centering\includegraphics[width=17cm]{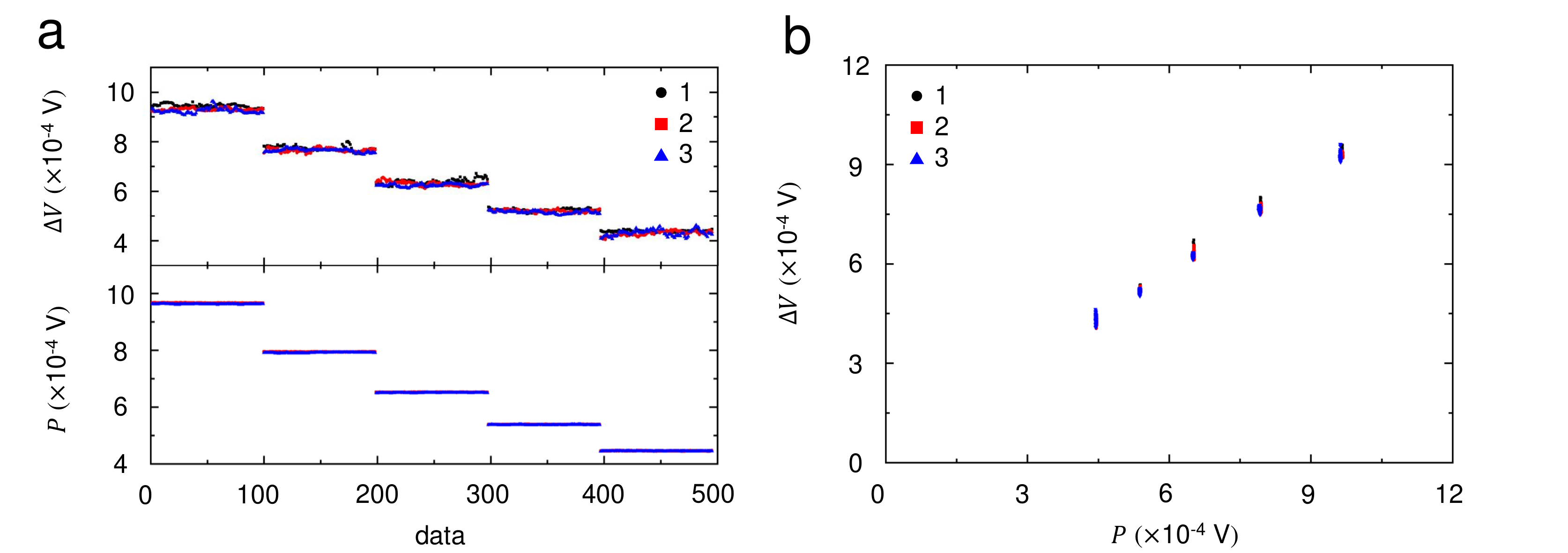}
\caption{\label{FigS5}Figure S6: (a) Measured probe signal response and lock-in magnitude of pump laser for three times of measurements for sample with $d=108.2$ nm and $r=200$ $\mu$m. (b) Probe reflectivity response with respect to lock-in magnitude of pump laser. Probe reflectivity response of calibration sample was averaged over three measurements.}
\end{figure}

\begin{figure}[h]
\centering\includegraphics[width=12cm]{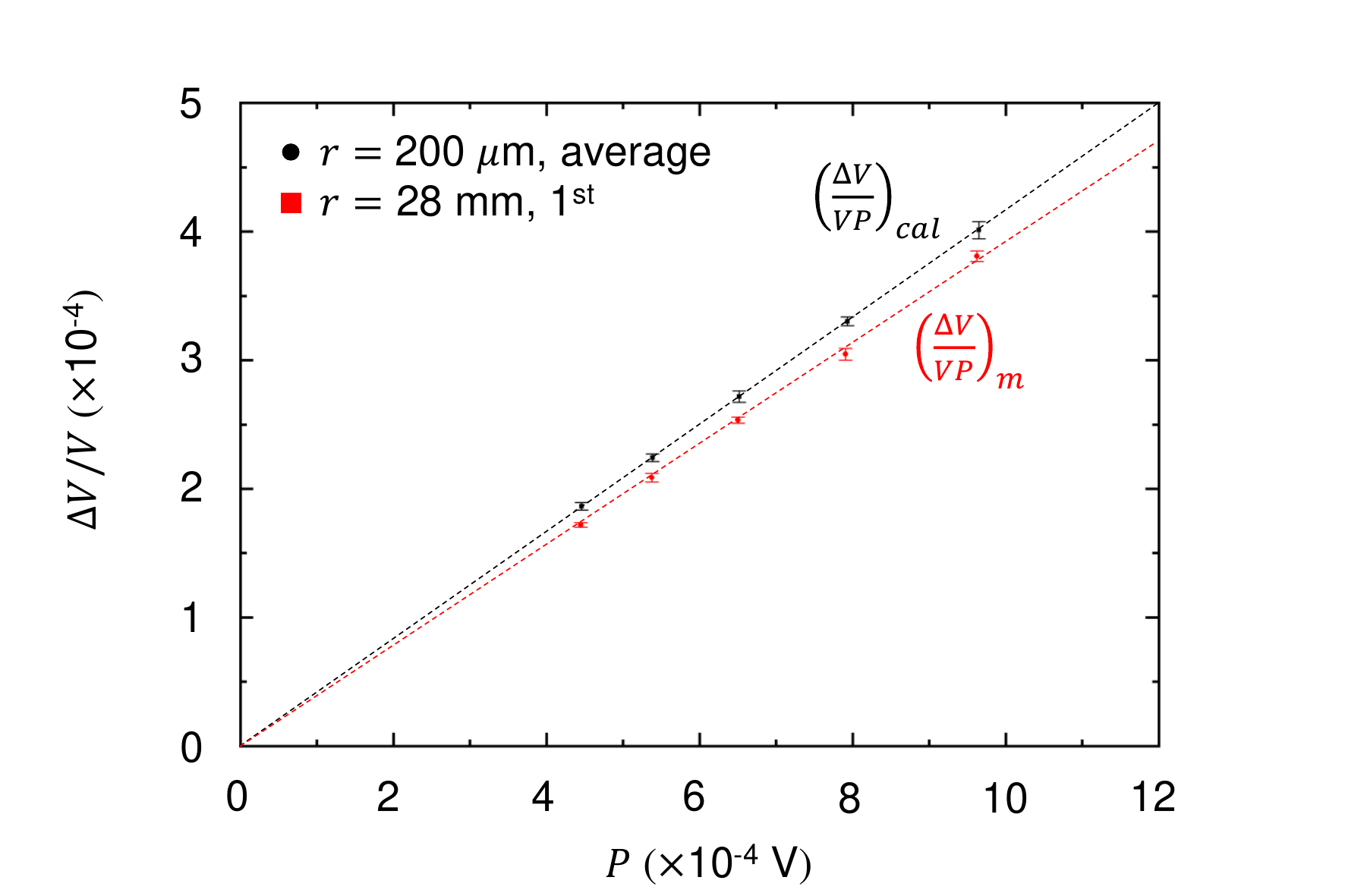}
\caption{\label{FigS6}Figure S7: Fitted results of $\frac{\Delta V}{V}$ with respect to $P$ for the calibration sample and the measurement sample.}
\end{figure}

The measurement error bar drawn in Fig. 2b of the manuscript represents the standard deviation of $k_{\parallel,spp}$ of three measurements for each Ti film. Here, we briefly show the fitting procedure of $k_{\parallel,spp}$ from experimental data. The probe signal response ($\Delta V$) with respect to lock-in magnitude of pump laser ($P$) was measured to obtain the $k_{\parallel,spp}$. The in-plane thermal conductivity of Ti film is fitted to $k_m$ by using measured data $\left( \frac{\Delta V}{VP} \right)_m$ and the following equation 
\begin{equation} \label{Eq:8}
	\left( \frac{\Delta V}{VP} \right) _m=\gamma \left( \frac{\Delta T(k_m)}{Q} \right)
\end{equation}
where temperature rise $\Delta T(k_m)$ with respect to heated power $Q$ is derived from 2-D heat diffusion model. In this equation, $\gamma$ can be obtained with a calibration sample (i.e., $r= 200$ $\mu$m) as 
\begin{equation} \label{Eq:9}
	\gamma=\left( \frac{\Delta T(k_{cal})}{Q} \right)_{cal}^{-1}\left( \frac{\Delta V}{VP} \right)_{cal}
\end{equation}
where $\left( \frac{\Delta V}{VP} \right)_{cal}$ is the measured data obtained with the calibration sample, and $k_{cal}$ is in-plane thermal conductivity of the calibration sample, which is same with $k_{\parallel,e}$ because the enhancement via surface plasmon polaritons is negligible when $r= 200$ $\mu$m. Measured $\Delta V$ and $P$ during the three measurements with calibration sample is shown in Fig. S\ref{FigS5}a. For the interval with consistent output power of the pump laser, $P$ and $\Delta V$ were measured for 5 seconds. Because  a datum was obtained every 50 ms, 100 data points were collected per interval. Besides, we measured $V$ by monitoring the one channel of a balanced photodetector and by modulating the probe laser before every measurement, which showed a deviation within 1\% with the average value of 2.31 mV. $\Delta V$ and $P$ are found to have standard deviations within 1.5\% and 0.1\%, respectively. Measured data $\Delta V$ with respect to $P$ for three measurements are shown in Fig. S\ref{FigS5}b and they are averaged to obtain the $\left( \Delta V/V\right)_{cal}$ with respect to $P$ plotted in Fig. S\ref{FigS6}, with standard deviation less than 1.5\%.  \\

%%%%%%%%%%%%%%%%%%%%
%%%%%%%%%%%%%%%%%%%%%%%
\newpage

%%%%%%%%%%
\begin{table}[!h]
	\caption{Table S4. Uncertainty analysis of $k_{\parallel,spp}$ for sample 1 with $r=28$ mm.}
	\label{Tab:4}
	\small
	\begin{center}
	\begin{tabular}{c| c |  c |  c|  c}
	\hline
	\hline
$k_{ref}$ & $\left( \frac{\Delta V}{VP} \right)_{cal, 95\%}$ &  $\gamma$   & $\left( \frac{\Delta V}{VP} \right)_{m,95\%}$  & $k_m$  \\
(W/m$\cdot$K) & (V$^{-1}$)  & ($\times10^{-5}$)  & (V$^{-1}$)  &(W/m$\cdot$K)  \\
\hline
7.14$\pm$0.15   & 0.4163$\pm$0.0008   & 1.431$\pm$0.003 & 0.3905$\pm$0.0051 & 2.8$\pm$0.6 \\
\hline
    \hline
\end{tabular}
\end{center}
\end{table}
%%%%%%%%%%%%

The fitting procedure for one-time measurement of Sample 1 with $r=28$ mm is summarized in Table S\ref{Tab:4}. From $k_{cal}$, $\frac{\Delta T(k_{cal})}{Q}$ can be obtained using the 2-D heat diffusion model. Also, $\left( \frac{\Delta V}{VP} \right)_{cal, 95\%}$ be derived by fitting 5 measured $\frac{\Delta V}{V}$ values with respect to $P$ for the calibration sample in the range of 95\% confidence limit. Then, $\gamma$ can be calculated from Eq.\ (5). Similarly, $\frac{\Delta V}{V}$ with respect to $P$ for the measurement sample ($r=28$ mm in this case), can be also fitted as $\left( \frac{\Delta V}{VP} \right)_{m,95\%}$ as shown in Table S\ref{Tab:4}. With the obtained $\gamma$, $\frac{\Delta T(k_{m})}{Q}$ can be also estimated, and the in-plane thermal conductivity of measurement sample ($k_m$) can be also determined based on the 2-D heat diffusion model. In Fig. S\ref{FigS6}, experimental results for the measurement sample (with subscript `m') and the calibration sample (with subscript `cal') are plotted. From the above fitting procedure, $k_m$ was derived as 2.8 W/m$\cdot$K with an uncertainty of 0.6 W/m$\cdot$K. The uncertainty can be reduced by repeating measurements according to the following equation:
\begin{equation} \label{Eq:10}
	u_{tot}=\sqrt{u_a^2+u_b^2} = \sqrt{u_a^2+\frac{1}{N^2}\sum^N_{k=1}[u_b(x_k)]^2}
\end{equation}
where $u_a$ is the Type A uncertainty derived from standard deviation of the data \cite{joint2008evaluation}, $N$ refers to the number of data in each measurement, and $u_b(x_k)$ is the measurement uncertainty for a single datum, $x_k$. In this case, $x_k$ is $\left( \frac{\Delta V}{VP} \right)$. By conducting three measurements, $u_b$ of the sample 1 with $r=28$ mm can be reduced to 0.36 W/m$\cdot$K, which is 15\% of its average value.  
%%%%%%%%%%%%%%%%%%%%
%%%%%%%%%%%%%%%%%%%%%%%

\newpage

\section*{8. Measurement of Ti film thickness with stylus profiler}

Three types of samples were fabricated by e-beam depositing Ti films with thicknesses of 108.2, 302.7, and 1002.7 nm on 4-inch glass wafers. The thicknesses of the Ti films were measured with a stylus profiler (Alpha-Step 500, KLA TENCOR CORP.). Scanning length on the film was set to be 100 $\mu$m, and scanning speed was 10 $\mu$m/s. Measurement results are summarized in Table S\ref{Tab:5}. The average thickness of Sample 1 is 108.2 nm and every sample deviates from its average within 2.3\%. Similarly, the average thicknesses of Sample 2 and Sample 3 are 302.7 nm and 1002.7 nm, respectively, and have standard deviations of 1.5\% and 3.6\% from their average values. Due to the uncertainty of the film thickness, the theoretical value of $k_{\parallel,spp}$ of each sample can fluctuate within 2\%, 1.5\%, and 3.5\% of the one obtained with the average thickness shown in Fig. 2b of the manuscript, respectively.\\

%%%%%%%%%%
\begin{table}[!h]
	\caption{Table S5. Measured Ti film thickness.}
	\label{Tab:5}
	\small
	\begin{center}
	\begin{tabular}{c| c |  c |  c}
	\hline
	\hline
	 & \multicolumn{3}{c}{d (nm)}  \\
	 \hline
r  (mm)   & Sample 1 & Sample 2 & Sample 3 \\
\hline
0.2 & 110      & 300      & 1060     \\
1      & 110      & 310      & 960      \\
1.5    & 110      & 310      & 1050     \\
2      & 110      & 300      & 1000     \\
3      & 110      & 300      & 1000     \\
4      & 110      & 300      & 980      \\
6      & 105      & 305      & 1000     \\
9      & 110      & 305      & 970      \\
12     & 105      & 305      & 1050     \\
20     & 105      & 295      & 960      \\
28     & 105      & 300      & 1000     \\
\hline
   average    & 108.2 & 302.7 & 1002.7 \\
    stdev.   & 2.5 & 4.7 & 36.1\\
    \hline
    \hline
\end{tabular}
\end{center}
\end{table}
%%%%%%%%%%%%

\clearpage

\section*{9. Spectral thermal conductivity of SPP}

\begin{figure}[h]
\centering\includegraphics[width=17cm]{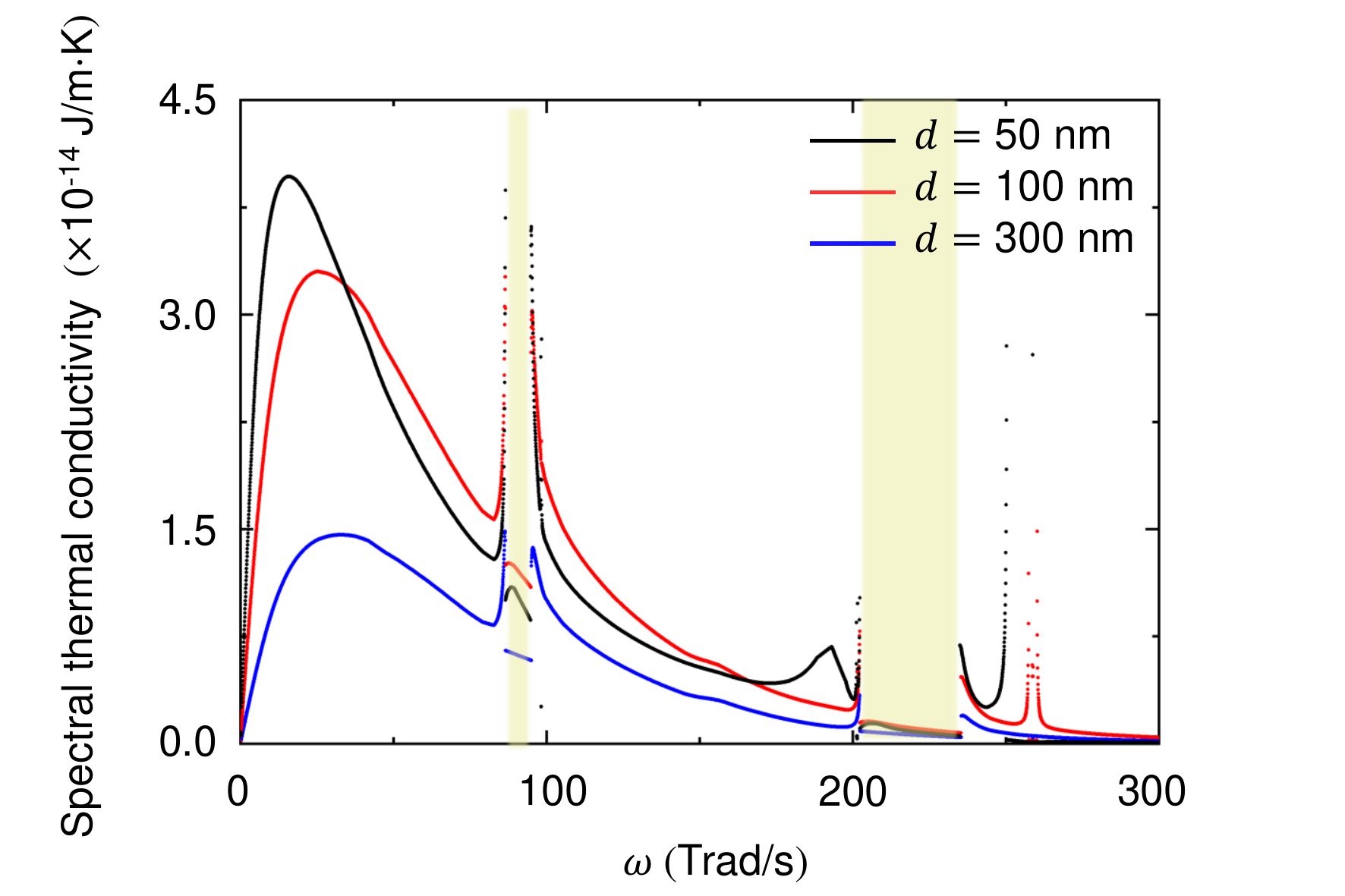}
\caption{\label{spectral}Figure S8: Spectral thermal conductivity of Ti film on glass substrate with three different film thicknesses when $L=28$ mm.}
\end{figure}

Spectral thermal conductivity (i.e., $k_\omega=1/(4\pi d)\hbar\omega\Lambda_{eff}\beta_R(df_0/dT)$) of Ti film on glass substrate for three different film thicknesses is shown in Fig.\ S\ref{spectral}. $L$, which is the maximum propagation length of the sample, is assumed to be 28 mm. The spectral thermal conductivity of the samples with three different film thicknesses has a maximum value at frequencies below 100 TRad/s. Even when $L$ is set to be less than 28 mm, the spectral thermal conductivity has substantial value only in the frequency range within 300 Trad/s.   \\

%%%%%%%%%%%%
%%%%%%%%%%%%%

\newpage

\section*{10. Dispersion curve near the light line of air}

\begin{figure}[h]
\centering\includegraphics[width=17cm]{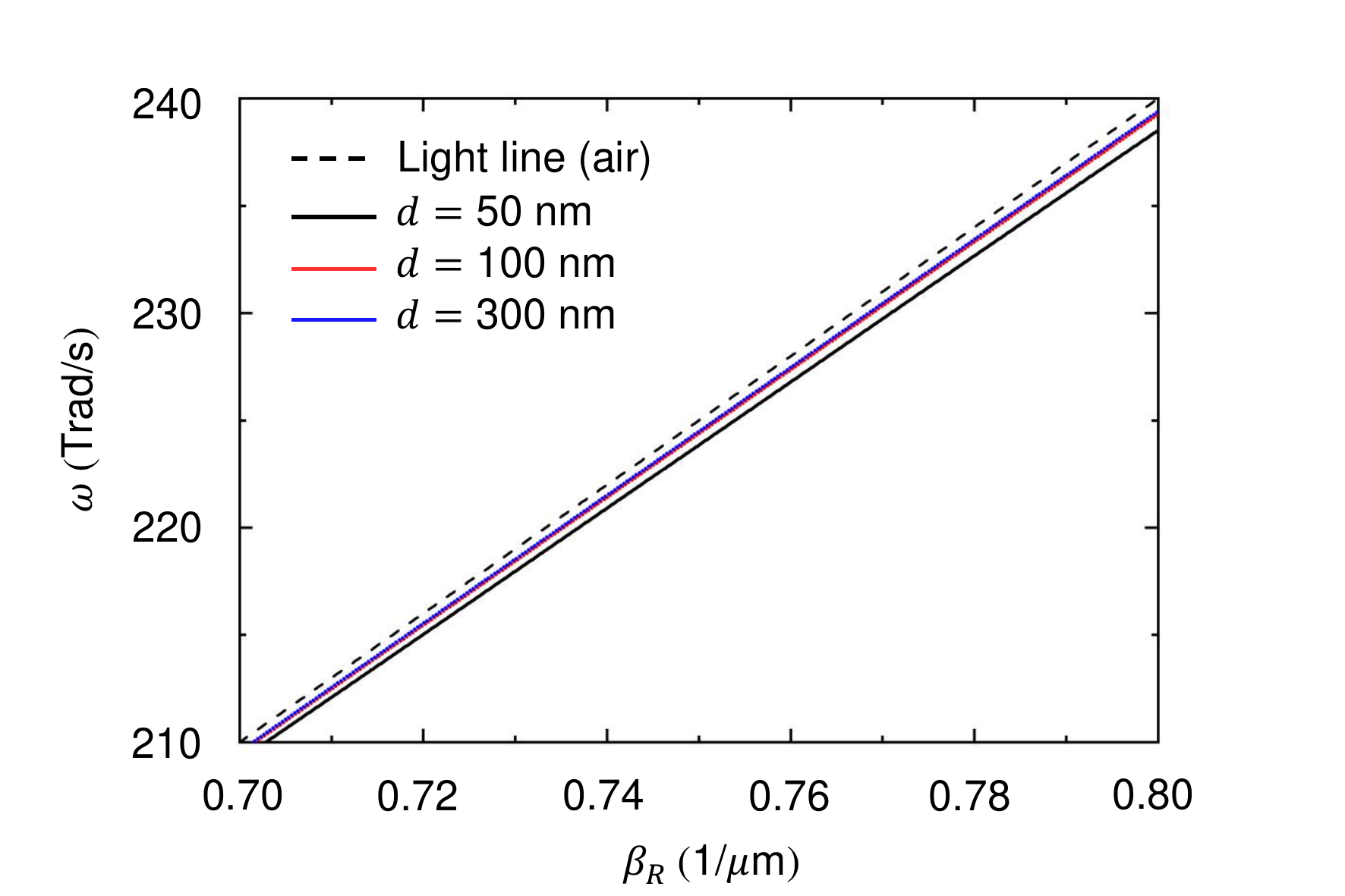}
\caption{\label{dispersion}Figure S9: Light line of air and dispersion curves of $\beta_R$ of SPPs supported by Ti film on glass substrate for three different Ti film thicknesses. Dispersion curve moves close to the light line of air as film thickness increases.}
\end{figure}

The dispersion curves of the real part of the in-plane wavevector $\beta_R$ for surface plasmon polaritons (SPPs) supported by Ti film on a glass substrate near the air light line for three different film thicknesses are shown in Fig.\ S\ref{dispersion}. The dispersion curves almost superimpose on the light line of air, thus showing photon-like behavior. In the enlarged figure in Fig.\ S\ref{dispersion}, it can be clearly noted that the dispersion curve becomes closer to the light line of air as the thickness of the film increases.  \\

%%%%%%%%%%%%
%%%%%%%%%%%%%

\newpage

\section*{11. Analytic solution of SPP}

\begin{figure}[h]
\centering\includegraphics[width=17cm]{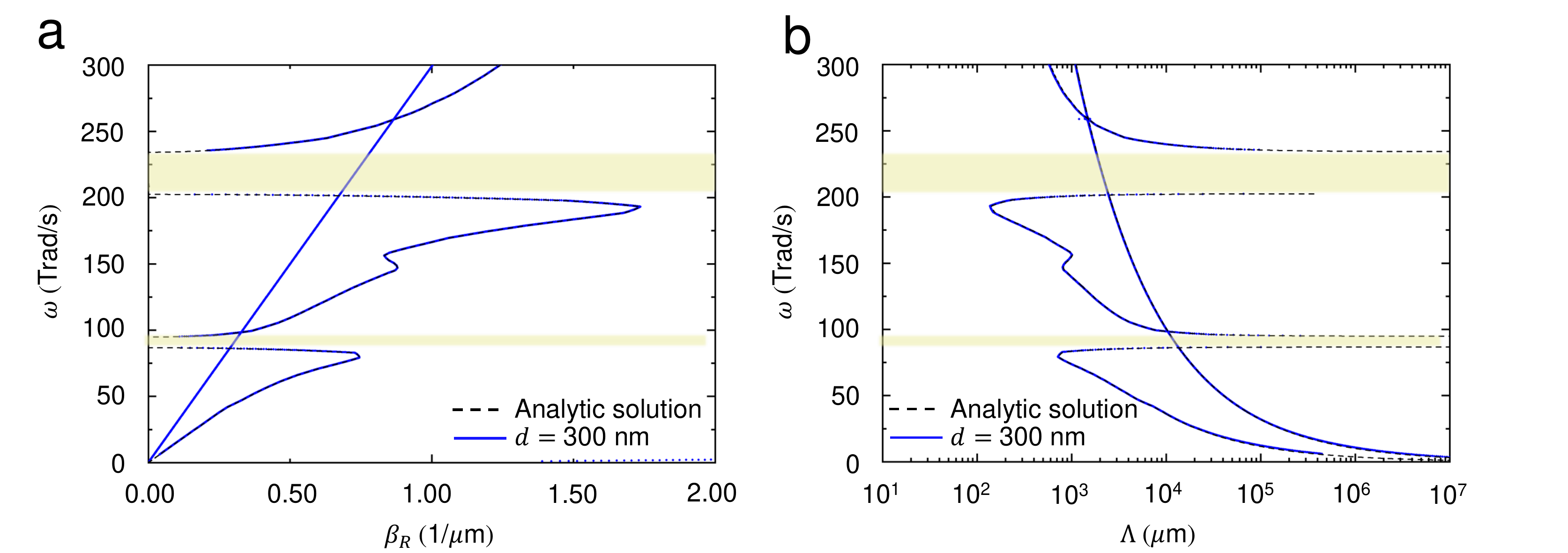}
\caption{\label{analytic}Figure S10: Numerical solution of (a) $\beta_R$ and (b) $\Lambda$ for Ti film on glass substrate when Ti film thickness is $d=300$ nm. Dashed curves showing analytic solutions of dispersions of SPPs at semi-infinite metal/dielectric interfaces (i.e., Ti/air interface and Ti/glass interface) are also overlaid.}
\end{figure}

The real part of dispersion $\beta_R$ and propagation length $\Lambda$ for SPPs supported by Ti film ($d=300$ nm) on a glass substrate are shown in Fig.\ S\ref{analytic}. Analytic solutions of dispersions for SPPs at a semi-infinite metal/dielectric interface are also overlaid, which can be defined as \cite{burke1986surface}
\begin{equation} \label{Eq:11}
	\beta=\varepsilon_d^{1/2}k_0 \left( \frac{\varepsilon_m}{\varepsilon_m + \varepsilon_d} \right)^{1/2}
\end{equation}
 \\
where $k_0$ is wave vector at vacuum, $\varepsilon_m$ is the dielectric function of metal and $\varepsilon_d$ is dielectric function of dielectric. Dashed curves shown in Fig. S\ref{analytic}a are readily calculated from Eq.\ \eqref{Eq:11} by letting metal as Ti and dielectrics as air and glass. To obtain the propagation length of SPPs ($\Lambda=1/2\beta_I$) at a semi-infinite metal/dielectric interface, we used an approximate analytic solution for the imaginary part of the in-plane wavevector $\beta_I$ for SPPs at a semi-infinite metal/dielectric interface as in \cite{burke1986surface}
\begin{equation} \label{Eq:12}
	\beta_I=\varepsilon_d^{3/2}k_0\varepsilon_I/2 |\varepsilon_m|^2
\end{equation}
 \\
 where $\varepsilon_I$ is the imaginary part of $\varepsilon_m$. Accordingly, dashed curves plotted in Fig. S\ref{analytic}b are calculated from Eq. \eqref{Eq:12}. Figure S\ref{analytic} clearly shows that SPPs supported by Ti film with a thickness of $d=300$ nm almost coincide with SPPs at each interface of Ti film (i.e., semi-infinite metal/dielectric interface), implying that SPPs at each interface of 300-nm-thick Ti film are decoupled. \\
  
%%%%%%%%%%%%
%%%%%%%%%%%%%

\newpage

\section*{12. Beam diameter of probe and pump laser}

\begin{figure}[h]
\centering\includegraphics[width=17cm]{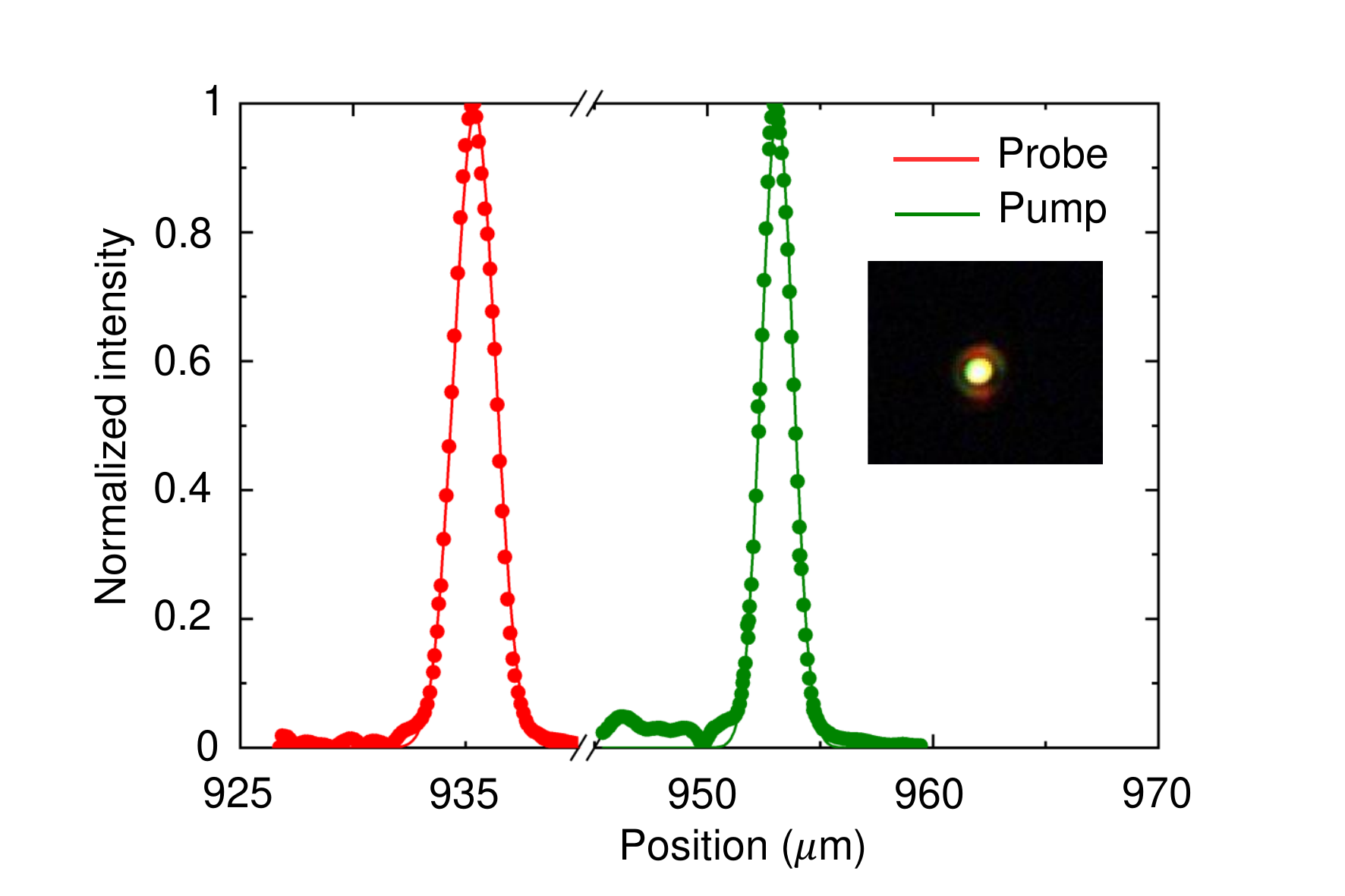}
\caption{\label{FigS9}Figure S11: Beam profile of the probe laser (red) and the pump laser (green). Inset shows the beam profile of probe and pump laser in CCD.}
\end{figure}

The beam size of the probe laser and pump laser passed through a 40X objective lens (UPLFLN 40X, Olympus) was measured with the knife-edge method \cite{yang2013thermal}. The gold pad that is deposited on the glass substrate moves in the $x$ direction with a motorized stage (Thorlabs Z8) while the reflectance of the beam is monitored. The beam profile can be obtained by measuring the reflectivity change (i.e., beam intensity) for position $x$, which varies as the beam deviates from the edge of the gold pad on the glass substrate. The Gaussian diameters ($1/e^2$) of the probe laser and the pump laser are 4.9 $\mu$m and 5 $\mu$m in the $x$ direction, while they are 4.6 $\mu$m and 4.3 $\mu$m in the $y$ direction. The beam profile on the sample was always monitored by CCD before every thermal conductivity measurement of the Ti film, as shown in the inset of Fig. S\ref{FigS9}.  \\

%%%%%%%%%%%%
%%%%%%%%%%%%%

\newpage

\section*{13. Thermal conductivity of glass substrate}

To derive the thermal conductivity of the glass wafer $k_s$, thermal diffusivity, heat capacity, and density were measured as shown in Table S\ref{Tab:6}. We measured the thermal diffusivity of a glass wafer (i.e., fused silica) using laser flash analysis (Netzsch LFA 457). Also, the heat capacity of the glass was estimated with differential scanning calorimetry (Netzsch DSC 204 F1 Phoenix), and the density was obtained by measuring the dimension and weight of the samples with vernier callipers (Mitutoyo) and an electronic scale (AND GX400). The resultant thermal conductivity of the glass wafer is determined as 1.35 W/m$\cdot$K. Given that uncertainties of LFA 457 and DSC 204 F1 are 0.3\% and 3\%, respectively and measurement uncertainty of the density is estimated as 0.01 g/cm$^3$ (uncertainties of the vernier calipers and the AND GX400 are 0.01 mm and 0.001 g, respectively), from the error propagation analysis, measurement uncertainties of the thermal conductivity of the fused silica can be obtained by \cite{joint2008evaluation}
\begin{equation} \label{Eq:1}
	u_c=\sqrt{\sum^n_i(c_i^2 u_i^2)}
\end{equation}
where $c_i$ is the sensitivity coefficient of the variables given as $c_i=\frac{\partial k_s}{\partial x_i}$, with $x_i$ being the variable (i.e., thermal diffusivity, heat capacity, density), and $u_i$ denotes the uncertainty of each variable, which can be estimated from the root sum square of the Type A and Type B uncertainties. Uncertainty of the glass thermal conductivity is within 3\% of its value. \\

%%%%%%%%%%
\begin{table}[!h]
	\caption{Table S6. Measured material properties for derivation of thermal conductivity of glass wafer}
	\label{Tab:6}
	\small
	\begin{center}
		\begin{tabular} {c| c |  c |  c |c   } 
		\hline\hline
		Material & Thermal diffusivity & Specific heat capacity & Density & Thermal conductivity\\
	             & (mm$^2$/s) & (J/g$\cdot$K) & (g/cm$^3$) & (W/m$\cdot$K)\\
		\hline
     	SiO$_2$ & $0.777\pm0.002$ & $0.779\pm0.02$ & $2.24\pm0.01$ & $1.35\pm0.04$ \\                     
   		\hline\hline    
		\end{tabular}
	\end{center}
\end{table} 
%%%%

\newpage

\bibliographystyle{naturemag}
\bibliography{Kim_supplemental_material}